\documentclass[10pt]{article}
\usepackage{amsmath,wasysym}

\usepackage{amssymb}

\usepackage{graphicx}
\usepackage[dvips]{epsfig}

\usepackage{cite}

\usepackage{color}

\usepackage[labelfont=bf,labelsep=period,justification=raggedright]{caption}

\usepackage{rotating}
\usepackage{mathrsfs}
\usepackage{pb-diagram}
\usepackage[all]{xy}
\usepackage{makeidx}
\usepackage{indentfirst}
\usepackage{multicol}
\usepackage{amsfonts}
\usepackage{epsfig}
\usepackage{subfigure}

\usepackage{cite} 
\usepackage{url}  
\usepackage{ifthen}  
\usepackage{multicol}   
\usepackage[utf8]{inputenc} 

\newboolean{publ}

\newcommand\bm[1]{{\protect\bf #1}}

\def\Y{\bm{y}}    
\def\X{\bm{x}}
\def\U{\bm{u}}

\def\H{\bm{h}}
\def\F{\bm{f}}
\def\G{\bm{g}}
\def\P{\bm{p}}

\def\includegraphics{}

\makeatletter
\renewcommand{\@biblabel}[1]{\quad#1.}
\makeatother

\date{}

\begin{document}

\begin{flushleft}
{\LARGE
\textbf{Sloppy models can be identifiable}
}
\\[0.5cm]
{\large Oana-Teodora Chis, 
Julio R. Banga, 
Eva Balsa-Canto$^{\ast}$}
\\[0.5cm]
Bioprocess Engineering Group,\ IIM-CSIC,\ Vigo, Spain
\\
$^{\ast}$ E-mail: ebalsa@iim.csic.es
\end{flushleft}

\section*{Abstract}

Dynamic models of biochemical networks typically consist of sets of non-linear ordinary differential equations involving states (concentrations or amounts of the components of the network) and parameters describing the reaction kinetics. 
Unfortunately, in most cases the parameters are completely unknown or only rough estimates of their values are available. Therefore, their values must be estimated from time-series experimental data. 

In recent years, it has been suggested that dynamic systems biology models are universally sloppy so their parameters cannot be uniquely estimated. In this work, we re-examine this concept, establishing links with the notions of identifiability and experimental design. Further, considering a set of examples, we address the following fundamental questions: i) is sloppiness inherent to model structure?; ii) is sloppiness influenced by experimental data or noise?; iii) does sloppiness mean that parameters cannot be identified?,
and iv) can sloppiness be modified by experimental design?

Our results indicate that sloppiness is not equivalent to lack of structural or practical identifiability (although they can be related), so sloppy models can be identifiable. Therefore, drawing conclusions about the possibility of estimating unique parameter values by sloppiness analysis can be misleading. Checking structural and practical identifiability analyses is a better approach to asses the uniqueness and confidence in parameter estimation.

\section*{Author Summary}
The model building cycle requires the reconciliation of the underlying hypothesis with experimental data. In the context of systems biology, this implies, in most of the cases, the necessity of identifying unknown kinetic parameters by data fitting. In this concern, recent works suggest that dynamic systems biology models are universally sloppy and, thus, parameters cannot be uniquely estimated. 

In this work we show how models regarded as sloppy are structurally identifiable thus, in principle, parameters can be given unique values. In the case of structural identifiability, it is only a matter of the experimental constraints and noise that the quality of the parameter estimates may be limited. In this sense, we analyze how sloppiness is affected by the experimental setup and experimental noise and we illustrate, with a number of examples related to biochemical networks, how sloppy models are indeed practically identifiable.

Our results indicate that sloppiness does not mean that parameters cannot be estimated and a complete identifiability analysis provides the tools to estimate ranges of parameters which are coherent with experimental data and can then be used to assess quality of predictions.

\section{INTRODUCTION}

Dynamic modeling of cellular processes has received great attention in the recent literature. The objective is to describe the interactions of distinct molecular entities (for example, proteins, transcripts or regulatory sites) which give rise to particular cellular behaviors. Typically the models consist of sets of non-linear ordinary differential equations involving a high number of states (concentrations or amounts of the components of the network) and a large number of parameters describing the reaction kinetics. 

Unfortunately, in most cases the parameters are completely unknown or only rough estimates of their values are available. Therefore, their values are usually estimated from time-series experimental data \cite{jaqaman-danuser:2006}. The so called parameter estimation problem is then formulated as an optimization problem where the objective is to find the parameter set so as to minimize a given cost function that relates model predictions and experimental data, e.g. the least squares function or a similar cost index. 

Typically, parameter estimation problems in dynamic models of biochemical systems are characterized by limited observability, large number of parameters and a limited amount of noisy data \cite{banga-balsa:08}. This means that the solution of the problem is in general challenging and, even when using robust and efficient optimization methods, computationally expensive. In addition, the usual limited amount of noisy data often results in great uncertainty on the values of the estimated parameters \cite{vanlier-tiemann-hilbers-vanriel:2013}.

In this regard, the works by Gutenkunst at al. \cite{gutenkunst-etal-1:2007, gutenkunst-etal:2007} suggest that dynamic systems biology models are universally sloppy and thus it is not possible to estimate unique values for the model parameters. The parameters of the models can be divided into stiff (those that can be determined with great certainty) and sloppy (those that can vary by orders of magnitude without influencing significantly the output of the model). The authors use the distribution of the eigenvalues of the Fisher information matrix as a means to assess sloppiness. Even though there is not clear cut-off between sloppy and stiff parameters, from the results published in the analysis of $17$ models, it may be concluded that a separation of more than $3$ orders of magnitude in the eigenvalues is enough to regard the model sloppy. 

The fact that the sloppiness can be a ``universal'' property of systems biology models has opened a debate on whether the modelers should focus on predictions instead of trying to estimate precise values for the parameters, which in the other hand is rather complicated or even ``impossible''. 

Many modelling works use the idea to justify the fact that several different values of the parameters provide the same fit (see for example, \cite{rand:2008,chen:2009,kotte:2009,wang:2009}, among many others) and thus parameters cannot be given unique values. However the origin of the problem is usually neither discussed nor sought.

Cedersund and coworkers \cite{cedersund-roll:2009,cedersund:2012} suggest the alternative of focusing on predictions. In this regard they have coined the term {\em core predictions} for those specific model properties that may be uniquely identified even in the case that parameter values may not. Recently they suggested an optimization based approach to compute the parameter regions complying with core predictions \cite{cedersund:2012}.  

In addition, the idea that a model being sloppy will be insensitive to changes on the parameters along sloppy directions, and highly sensitive along stiff combinations of parameters has been exploited to propose reduced order modelling techniques \cite{sunnaker-cedersund-jirstrand:2011,apri-deGee-molenaar:2012,berthoumieux-brilli-kahn-deJong-cinquemani:2012}. Eventually, a non-sloppy model will be obtained that is able to capture the relevant dynamics of the original model but with a reduced number of states and parameters. In the same line, the recent work by Machta and co-workers \cite{Machta-Chachra-Transtrum-Sethna:2013} suggests that both physics and sloppy models show weak dependence of macroscopic observables on microscopic details and allow effective descriptions with reduced dimensionality.

Alternative works suggested that sloppiness could be assimilated to the concept of poor or lack of identifiability \cite{raue:2011,cirit:2012} which have been present in the systems and control literature for decades  \cite{ljung-glad:94,walter-pronzato:96}. Other authors have suggested the use of model based experimental design techniques to reduce sloppiness \cite{apgar:2010,Liepe-Filippi-Komorowski-Stumpf:2013}. 

However the origin of sloppiness or the actual consequences of a model being sloppy have not been explored in detail. In this work, we address the following issues: 
\begin{itemize}
\item	Is sloppiness inherent to model structure? 
\item Is sloppiness influenced by experimental data or noise? 
\item Does sloppiness necessarily mean that parameters cannot be identified?  \item May sloppiness be modified by experimental design?
\end{itemize}
 
\section{METHODS}

We assume the model to be of the general form: 
\begin{equation} 
\Sigma(\P): \, \left\{%
\begin{array}{ll}
\dot{\X}=\F(\X,\P,\U;t,\X_0), \\ 

\Y=\bm{h}(\X,\P,\U;t,\X_0), \, \X(t_0)=\X_0(\P), \label{eq:model}
\end{array}%
\right.
\end{equation}

\noindent where $\X=(x_1,...,x_{n_x}) \in \textbf{R}^{n_x}$ is the state vector, $\U=(u_1,...,u_{n_u})\in \textbf{R}^{n_u}$ a $n_u-$dimensional input (control) vector, and $\Y=(y_1,...,y_{n_y})\in \textbf{R}^{n_y}$ is the $n_y-$dimensional output (experimentally observed quantities).  The vector of unknown parameters is denoted by $\P=(p_1,...,p_{n_p})\in \textbf{P},$ and in general is assumed to belong to an open and connected subset of $\textbf{R}^{n_p}.$ The entries of $\F$ and $\bm{h}$ are analytic functions of their arguments. These functions and the initial conditions may depend on the parameter vector $\P\in \textbf{P}.$ 

\subsection{Structural identifiability}

Structural identifiability regards the possibility of giving unique values to model unknown parameters from the available observables, assuming perfect experimental data (i.e. noise-free and continuous in time) \cite{walter-pronzato:97}. A parameter $p_i,$ $i=1,...,n_p$ is \emph{structurally globally (or uniquely) identifiable} if for almost any $\P^*\in \textbf{P}$, $\Sigma(\P)=\Sigma(\P^*) \Rightarrow p_i=p_i^*$, whereas a parameter $p_i,$ $i=1,...,n_p$ is \emph{structurally locally identifiable} if for almost any $\P^*\in \textbf{P}$ there exists a neighborhood $\textbf{V}(\P^*)$ such that
$\P\in \textbf{V}(\P^*)$ and $\Sigma(\P)=\Sigma(\P^*) \Rightarrow p_i=p_i^*$.

The recent work by Chis et al. \cite{chis-banga-balsa-canto-plos:2011} reviews the different alternative methods to perform the structural identifiability analysis for nonlinear models concluding that not universally valid method exists but that a combination of the generating series approach with identifiability {\it tableaus} is suitable for typically large scale dynamic models in systems biology. This method is implemented in the MATLAB based GenSSI toolbox  \cite{chis-banga-balsa-canto:2011} which is used in this work.

The underlying idea of the generating series approach is that the observables can be expanded in series with respect to time and inputs around a given time point ($t_0$), and that the uniqueness of the series coefficients guarantees the structural identifiability of the model. The series coefficients are computed by means of successive Lie derivative of $\H$ along the vector fields $\F$ and $\G$. The identifiability {\it tableaus} correspond to the Jacobian of the Lie derivatives with respect to the model parameters and help to decide on global or local structural identifiability of the model \cite{balsa-alonso-banga:2010}.

\subsection{Parameter estimation problem}

The above representation (\ref{eq:model}) is a sufficiently accurate mathematical description of the real system, i.e. the only uncertainty is represented by the vector of unknown parameters. This means that, provided the model is structurally identifiable, there is a unique ``true'' value of the parameters, denoted by $\P^*=(p_1^*,...,p_{n_p}^*)$, which make the model to represent a given data set and to predict the system behavior. This vector $\P^*$ is computed by means of data fitting, i.e. by solving an optimization problem devoted to minimize the log-likelihood function which for the case of Gaussian experimental noise reads:

\begin{equation}
\chi^2(\P)=\sum_{e=1}^{n_e}\sum_{o=1}^{n_y}\sum_{s=1}^{n_s}\frac{[\Y_{e,o,s}(\P,t_s)-\tilde\Y_{e,o,s}]^2}{\sigma_{e,o,s}^2}, \label{eq:llk}
\end{equation}

\noindent where $n_e$ is the number of experiments, $n_y$ the number of observables for each experiment and $n_s$ corresponds to the number of sampling times; $\Y_{e,o,s}(\P,t_s)$ denotes the output of the model (\ref{eq:model}) for the sampling time $t_s$ under the experimental conditions $e$ and $\tilde\Y_{e,o,s}$ is the corresponding experimental data; $\sigma_{e,o,s}^2$ regards the variance of the measurement noise.

\subsection{Sloppiness analysis}

Parameter sloppiness is assessed in terms of the eigenvalues and the eigenvectors of the following modified Fisher information matrix \cite{gutenkunst-etal:2007} as evaluated in the optimal value of the parameters $\P^*$:

\begin{equation}
F_{i,j}=\frac{1}{n_d}\sum_{e=1}^{n_e}\sum_{o=1}^{n_y}\sum_{s=1}^{n_s}\frac{1}{\sigma_{e,o,s}^2}\frac{\partial\Y_{e,o,s}}{\partial log(\P_i)}\frac{\partial\Y_{e,o,s}}{\partial log(\P_j)}, \label{eq:modFIM}
\end{equation}

\noindent where $n_d$ is the total amount of data, logarithms of the parameters are used to avoid scaling issues due to different orders of magnitude of the parameters.

The largest eigenvalues correspond to the so called ``stiff'' parameters and the smallest correspond to the ``sloppy'' parameters \cite{brown:2003}. Models are said to be sloppy when the maximum eigenvalue ($\lambda_{max}$) is orders of magnitude larger than the minimum eigenvalue ($\lambda_{min}$). Even if there is no clear cutoff between sloppy and stiff parameters from previous works \cite{waterfall:2006,gutenkunst-etal:2007} it is concluded that ${\cal C}_F=\frac{\lambda_{min}}{\lambda_{max}}\apprle 10^{-3}$ would correspond to a sloppy model.

\subsection{Practical identifiability analysis}
In order to assess the quality of the parameter estimates, several possibilities exist. Bootstrap or jack-knife approaches allow to compute robust confidence intervals.  However, the associated computational cost makes it difficult to use these methods for large scale models. Alternatively, the confidence interval ($\rho_i$) of  $\boldsymbol\theta^*_i$ may be obtained through the covariance matrix:

\begin{equation}
\pm t_{\alpha/2}^{\gamma}\sqrt{\bf{C}_{ii}}  \label{eq:confidence-interval}
\end{equation}
\noindent where $t_{\alpha/2}^{\gamma}$ is given by Students t-distribution, $\gamma=N_d-\eta$ degrees of freedom and $(1-\alpha)100\%$ is the confidence interval selected, typically $95\%$.

For non-linear models, there is no exact way to obtain $\bf C$. Therefore, the use of approximations has been suggested. Possibly the most widely used is based on the Cramm\`er-Rao inequality which establishes, under certain assumptions on the number of data and non-linear character of the model, that the covariance matrix may be approximated by the inverse of the Fisher information matrix (${\cal F}$) in its typical definition:

\begin{equation}
{\cal F}=E(\Big[\frac{\partial \chi^2(\P)}{\partial \P}\Big]^T\Big[\frac{\partial \chi^2(\P)}{\partial \P}\Big]), \label{eq:FIM}
\end{equation}

From the covariance matrix it is also possible to compute the correlation matrix:

\begin{eqnarray}
Cr_{ij}=\frac{C_{ij}}{\sqrt{C_{ii}}C_{jj}} & i=1,\ldots,n_p & j=1,\ldots,n_p
\end{eqnarray}

\noindent in such a way that two parameters $(p_i,p_j)$ are completely uncorrelated

\subsection{Role of the experimental error}

To analyze the role of the amount of experimental noise we have generated a significant number of realizations of noisy pseudo-data by adding noise to the output of the models. More precisely, we have considered the following case of Gaussian noise with known time varying variance: 
           
\begin{eqnarray}
\tilde\Y_{e,o,s}=\Y_{e,o,s}+ \epsilon_{e,o,s} & with & \epsilon_{e,o,s}=\sigma \times randC_{e,o,s} \times max(\Y_{e,o}) 
\end{eqnarray}

\noindent where $ \epsilon_{e,o,s}$ are normally distributed independent random variables with standard deviation $\sigma \times max(\Y_{e,o})$ and $randC_{e,o,s}$ is a random number drawn from the standard normal distribution $\cal{N}(0,1)$ defined for every experiment, observable and sampling time. Three scenarios were analyzed: $\sigma=0.05$, $\sigma=0.1$ and $\sigma=0.2$. 

Subsequently the parameter estimation problem was solved for each of the ($200$ in this work) pseudo-data realizations. The sloppiness (${\cal C}_F$) and the mean value of the confidence intervals ($\rho$) are then computed for each of the optimal solutions achieved.

\subsection{Optimal experimental design}

In order to improve the quality of parameter estimates it is possible to use the model to define new experiments. The idea is to formulate a dynamic optimization problem where the objective is to find those experimental conditions which result in maximum information content as measured by, for example, the Fisher information matrix, subject to the system dynamics Eqn.~(\ref{eq:model}) plus experimental constraints. The problem can be solved by a combination of the control vector parameterization (CVP) method and a suitable optimizer enabling the simultaneous design of several dynamic experiments with optimal sampling times \cite{balsa-alonso-banga:2008,balsa-alonso-banga:2010}. 

It should be noted that different ${\cal F}$ based criteria can be used for the purpose of optimal experimental design. The most widely used are the D-optimum and the E-optimum which correspond to the maximization of the determinant of ${\cal F}$ and the maximization of its minimum eigenvalue, respectively. Here we will also explore the minimization of the sloppiness, i.e. the minimization of ${\cal C}_F$. This type of design will be regarded as S-optimum. 

AMIGO (Advanced Model Identification using Global optimization)\cite{balsa-banga:2011}, a MATLAB based toolbox that covers all model identification steps, was used in this work for the purpose of parameter estimation, sensitivity and practical identifiability analysis and optimal experimental design.

\subsection{Models}

To perform the different analyses proposed in this work we have selected a collection of examples representative of biological systems and processes. The idea was to cover different sizes and types of non-linear terms such as general mass action, Michaelis-Menten or Hill kinetics. All selected models are regarded as sloppy.

\begin{itemize}

\item Three models from the BioModels database \cite{novere:2006}

\begin{itemize}

\item {\bf Example 1: A minimal model for the mitotic oscillator} \cite{goldbeter:1991}. The minimal model for the mitotic oscillator is based on the cascade of post-translational modification that modulates the activity of cdc2 kinase during the cell cycle. The dynamics is characterized by three state variables, $13$ parameters. All state variables can be observed.

\item {\bf Example 2: The genetic MAPK cascade}\cite{kholodenko:2000}. Eukaryotic signal transductions are widely characterized by the mitogen-activated protein kinase (MAPK) cascades. The MAPK cascades relay extracellular stimuli from the plasma membrane to targets in the cytoplasm and nucleus, initiating diverse responses involving cell growth, mitogenesis, differentiation and stress responses in mammalian cells. The system dynamics is represented by $8$ differential equations with $22$ parameters including a Hill coefficient.

\item {\bf Example 3: The growth-factor-signaling network in PC12 cells} \cite{brown-hill-calero-myers-lee-sethna-cerione:2004}. The model represents the actions of neuronal growth factor (NGF) and mitogenic epidermal growth factor (EGF) in rat pheochromocytoma (PC12) cells. These growth factors stimulate extracellular regulated kinase (Erk) phosphorylation using intermediate signaling proteins, with distinct dynamical profiles. The model consist of $28$ non-linear ordinary differential equations with $48$ parameters. All states are amenable to observation.
\end{itemize}

Mathematical formulations of Examples 1 to 3 can be found in the Supplementary file.

\item{\bf Example 4: Fed-batch reactor for ethanol production}\cite{hong:86}. This model encodes the ethanol production using the anaerobic fermentation of glucose with the help of \textit{Sacharomyces cerevisiae}. The unsteady state material balances are the following:
\begin{equation}\label{batch}
\begin{array}{ll}
\dot{x}=x\mu-u\frac{x}{V},\\
\dot{S}=-\nu_0x\mu+u\frac{S_0-S}{V},\\
\dot{P}=x\epsilon-u\frac{P}{V},\\
\dot{V}=u,\\
\end{array}
\end{equation}
\noindent where $x,S,P$ are the cell mass, substrate, and ethanol concentrations (g/L), $u$ corresponds to glucose concentration (g/L),  $V$ is the volume (L), $\mu=\frac{\mu_0}{1+\frac{P}{K_P}}\frac{S}{K_S+S}$ is the specific growth rate and $\epsilon=\frac{1}{1+\frac{P}{K_P^{'}}}\frac{S}{K_S^{'}+S}$ the specific productivity. The value of the parameters is $\mu_0^*=0.408,K_P^*=16,K_S^*=0.22,\nu_0^*=10,K_P^{'*}=71.5,K_S^{'*}=0.44$. Initial conditions are $[x=1, S=150, P=0, V=10]$.

\item {\bf Example 5: A linear biochemical pathway with $14$ steps}. The system, that could describe a linear biochemical network, is represented by fourteen differential equations, sixteen parameters, as follows:

\begin{equation}\label{cs5}
\begin{array}{ll}
\dot{x}_1=-\frac{v_m x_1}{(k_m+x1)}+p_1u, \\
\dot{x}_2=p_1x_1-p_2x_2, \\
\dot{x}_3=p_2x_2-p_3x_3, \\
\dot{x}_4=p_3x_3-p_4x_4,\\
\dot{x}_5=p_4x_4-p_5x_5,\\
\dot{x}_6=p_5x_5-p_6x_6, \\
\dot{x}_7=p_6x_6-p_7x_7,\\
\dot{x}_8=p_7x_7-p_8x_8,\\
\dot{x}_9=p_8x_8-p_9x_9,\\
\dot{x}_{10}=p_9x_9-p_{10}x_{10},\\
\dot{x}_{11}=p_{10}x_{10}-p_{11}x_{11},\\
\dot{x}_{12}=p_{11}x_{11}-p_{12}x_{12},\\
\dot{x}_{13}=p_{12}x_{12}-p_{13}x_{13},\\
\dot{x}_{14}=p_{13}x_{13}-p_{14}x_{14},
\end{array}
\end{equation}

\noindent where $x_i$ stands for the components concentrations. The initial conditions are assumed to be zero for all components but $x_1(0)=2$ and nominal parameter values are: $vm=1.6$, $km=2.8$, $p1=0.5$, $p2=1.0$, $p3=1.6$, $p4=1.2$, $p5=1.2$, $p6=0.3$, $p7=1.3$, $p8=0.8$, $p9=1.4$, $p10=1.5$, $p11=1.0$, $p12=1.8$, $p13=1.2$, $p14=0.4$.

\item {\bf Example 6: Six-gene regulatory network}. The system, as proposed in the DREAM6 parameter estimation challenge\footnote{http://www.the-dream-project.org/challenges/dream6-estimation-model-parameters-challenge}, assumes linear kinetics for mRNA degradation and protein synthesis (translation) and degradation. The mRNA synthesis follows a Hill-type kinetics unless for some cases where there is no regulatory input to a gene. The mathematical model reads:

\begin{equation}\label{cs6}
\begin{array}{ll}
                    \dot{mr_1}=pst_1 - mdr_1 mr_1,\\
                    \dot{p_1}= rst_1 mr_1 - pdr p_1,\\
                    \dot{mr_2}=pst_2 as_1 rs_1 - mdr_2 mr_2,\\
                    \dot{p_2}=rst_2 mr_2 - pdr p_2,\\
                    \dot{mr_3}=pst_3 as_3 rs_4 - mdr_3 mr_3,\\
                    \dot{p_3}= rst_3 mr_3 - pdr p_3,\\
                    \dot{mr_4}= pst_4 as_2  rs_2 - mdr_4 mr_4,\\
                    \dot{p_4}=rst_4 mr_4 - pdr p_4,\\
                    \dot{mr_5}=pst_5 rs_3 - mdr_5 mr_5,\\
                    \dot{p_5}= rst_5 mr_5 - pdr p_5,\\
                    \dot{mr_6}=pst_6 rs_5 - mdr_6 mr_6,\\
                    \dot{p_6}= rst_6 mr_6 -pdr p_6,\\
                    as_1 = \frac{(p1/k_2)^{h_2}}{1+(p1/k_2)^{h_2}};  as_2 = \frac{(p1/k_2)^{h_1}}{1+(p1/k_1)^{h_1}}; 
                    as_3 = \frac{(p1/k_3)^{h_3}}{1+(p1/k_3)^{h_3})},\\
                    rs_1 = \frac{1}{1+(p6/k_5)^{h_5}};
                    rs_2 = \frac{1}{1+(p5/k_8)^{h_8}};
                    rs_3 = \frac{1}{1+(p4/k_6)^{h_6}},\\
                    rs_4 = \frac{1}{1+(p2/k_4)^{h_4}};
                    rs_5 = \frac{1}{1+(p4/k_7)^{h_7}}
\end{array}
\end{equation}

\noindent where $p_i$ represent proteins,$mr_i$ represent mRNA, $pdr$ represents the protein degradation constant, $mdr_i$ correspond to the mRNA degradation rate, $pst_i$ are the corresponding promoter strengths, $rs_i$ the strengths of the ribosomal binding sites, $k_i$ are the binding affinities and $h_i$ correspond to the Hill coefficients. All mRNA degradation rate constants are assumed to be known and equal to $1$. All other parameters are to be estimated. In this work, selected nominal values are: $pdr=0.8$, $pst_1=pst_5=pst_6=3$, $pst_2=4.7$, $pst_3=5$, $pst_4=20$, $rst_1=1$, $rst_2=rst_3=rst_6=4$, $rst_4=0.4$, $rst_5=6$, $k_1=k_5=1$, $k_2=0.9$, $k_3=k_6=k_7=0.1$, $k_4=9.5$, $k_8=0.2$,  $h_1=h_4=h_8=4$, $h_2=h_3=h_6=h_7=2$, $h_5=1$. At initial time all the protein concentrations are set to $1$ and those corresponding to mRNAs are set to $0$.    
\end{itemize}

\section{RESULTS}

\subsection{Sloppiness {\it versus} structural identifiability}

The structural identifiability analysis was performed to all examples, concluding that all of them are at least locally structurally identifiable. Figures 1a-c present the identifiability {\em tableaus} for the Examples 1-3 selected from the BioModels database. Model formulations and further details on the identifiability tests can be found in the Supplementary info.

\begin{center}
\epsffile{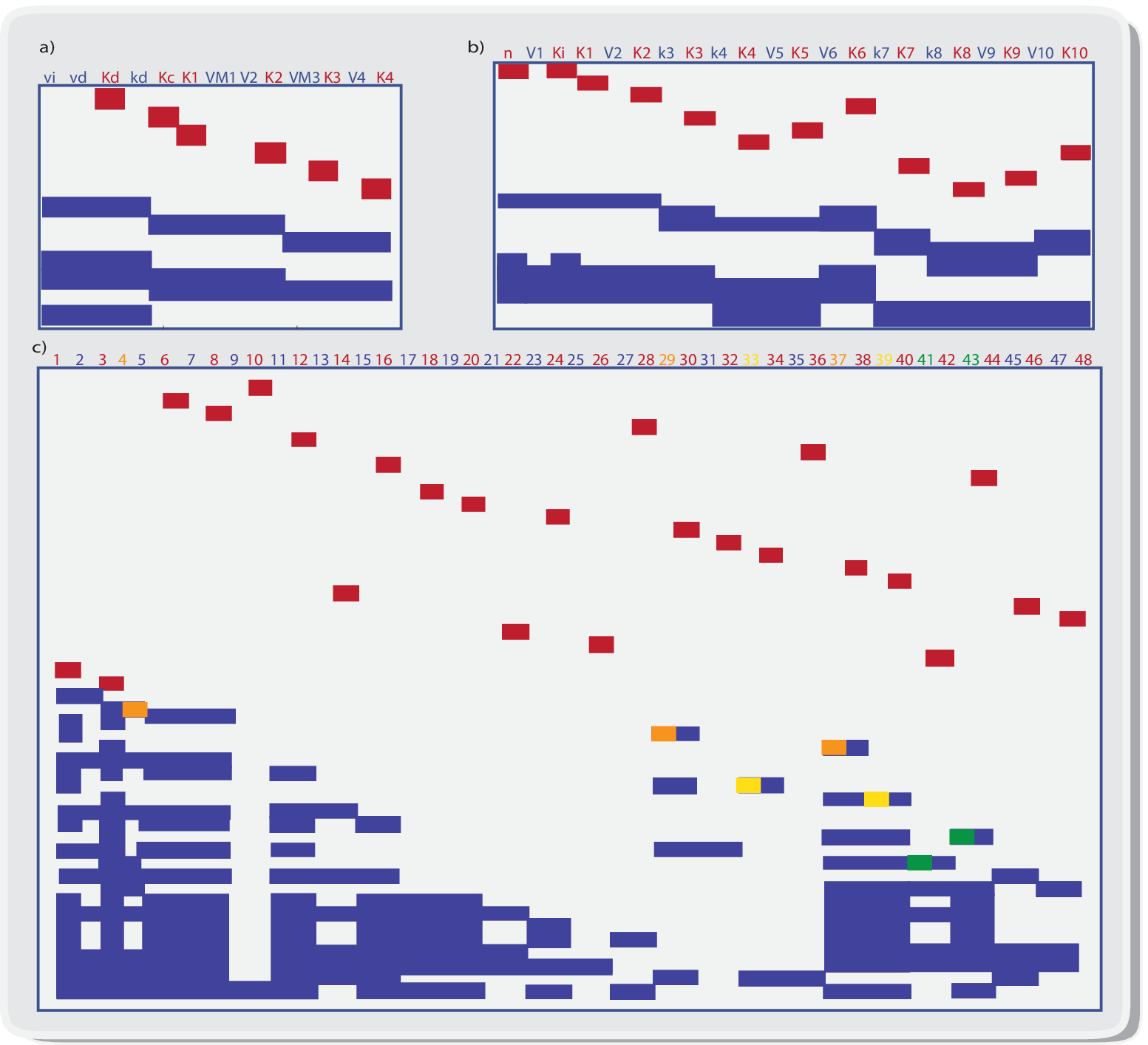} \\
\end{center}
{\bf Figure 1. Identifiability {\em tableaus}.} {\small a) Example A minimal model for the mitotic oscillator, b) the genetic MAPK cascade and  c) the growth-factor-signaling network. In red, direct globally identifiable parameters; in orange, yellow and green, $2nd$ to $4th$ levels of globally identifiable parameters; in blue, at least locally identifiable parameters.}

\subsection{Sloppiness {\it versus} practical identifiability}

To analyze the relationship between practical identifiability and sloppiness we considered examples 4 to 6 under different experimental setups. 

\subsubsection*{Example 4}
For the Example 4, related to the production of ethanol in a fed-batch reactor, we considered the scenario in which the reactor is fed with a constant amount of glucose $u=10$ for $t_f=24 h$ and $30$ (Case 1a) or $60$ (Case 1b) samples are taken for $x, S, P$ equidistantly distributed throughout the experiment duration. This corresponds to a total amount of $90$ data in Case 1a and $180$ data in Case 1b. 

Figure 2 presents an overview of the results. Figure 2a accounts for the sloppiness of the system. Results show how the system is sloppy in both scenarios. However the distribution of eigenvalues and the values of ${\cal C_F}$ change when adding sampling times (${\cal C_F}=1\times 10^{-9}$ and ${\cal C_F}=4.3\times 10^{-7}$). Figure 2b shows the corresponding confidence regions revealing that the addition of sampling times increases the magnitude of ${\cal F}$, thus improving identifiability. It should be noted, however, that confidence intervals for $K_s$ and $K'_s$ are still rather large thus indicating that the model is poorly practically identifiable. Figure 2c shows the correlation matrix for both scenarios. Results reveal that under constant feeding parameters $K_p$, $K_s$, $K'_s$ and $K'_p$ are rather correlated, being $K'_s$ and $K_s$ highly correlated in both cases. Figures 2d present model predictions for different $K'_s$ values while keeping all other parameters fixed to the optimum. As it can be seen the cell mass and the substrate are insensitive to changes in $K'_s$ whereas ethanol production is affected only at the end of the experiment. As a result the fact that a parameter is sloppy does not necessarily mean that model predictions will remain the same independently of the parameter value. Note however that since $K'_s$ and $K_s$ are highly correlated changes in $K'_s$ can be easily compensated with changes in $K_s$ and predictions will remain unchanged if both parameters are appropriately manipulated.

\begin{center}
\epsffile{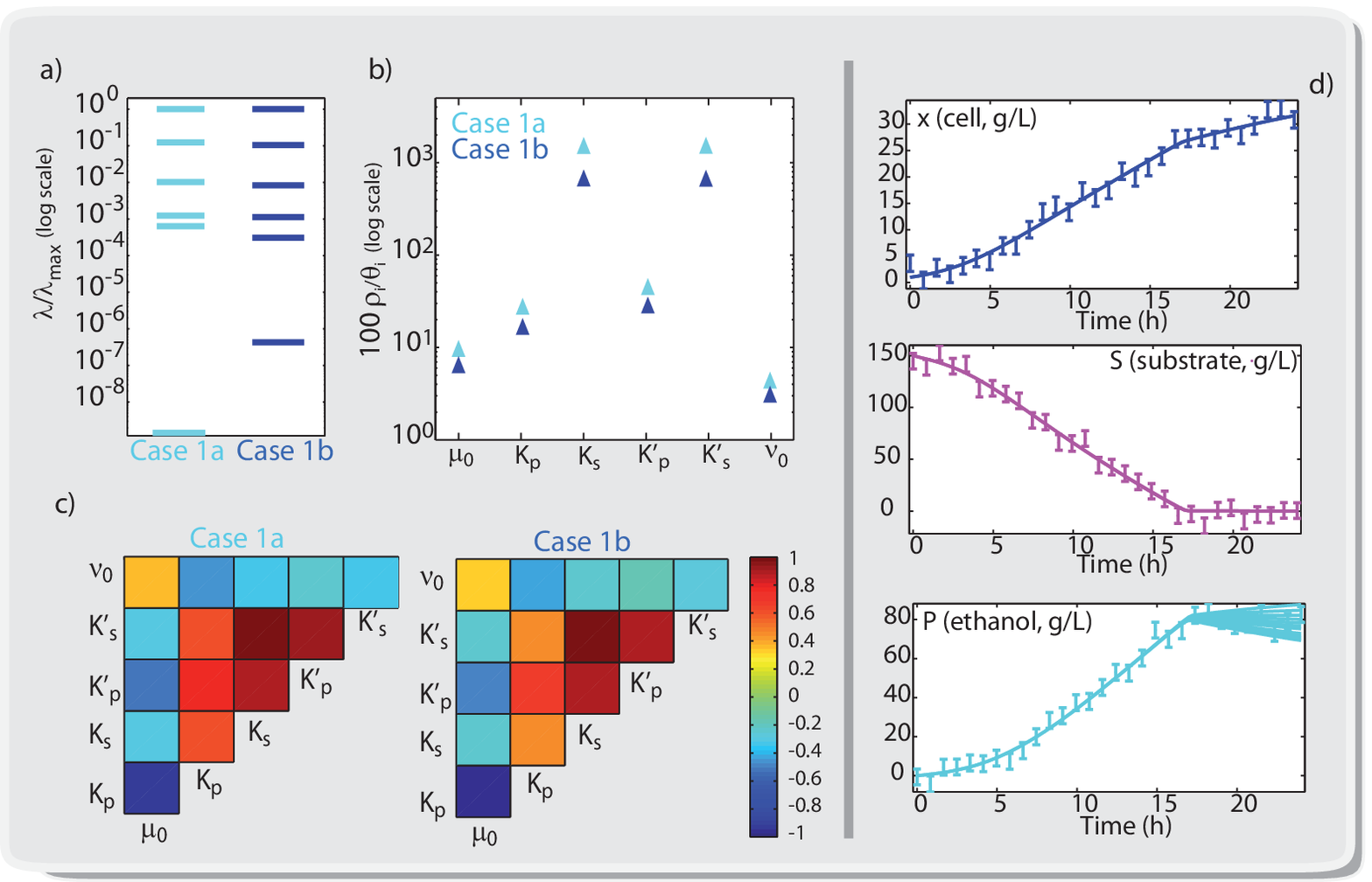} \\
\end{center}
{\bf Figure 2. Sloppiness vs practical identifiability for the ethanol production example.} {\small a) Normalized eigenvalue spectra as a measure of the sloppiness for the different scenarios. The eigenvalues span between 7 and 9 orders of magnitude, indicating sloppiness in both cases. b) Normalized confidence intervals for the different model parameters for both scenarios. c) Correlation matrices. d) Model predictions when modifying the value of $K'_s$ between $0.5$ and $2$ times its optimal value (all other parameters are fixed to the optimum). Plots suggest that $x$ and $P$ are not sensitive to $K'_s$ and therefore those measurements are not providing information to estimate its value, at least for the selected constant feeding profile.}

It should be also noted that only last samples of ethanol are the ones providing information to estimate $K_s$ and $K'_s$. Therefore it is expected that the accumulation of sampling times at the end of the experiment will improve identifiability. 
In fact the use of $30$ non equidistant sampling times (Case 1c) reduces the sloppiness as compared to the use of $60$ equidistant sampling times (${\cal C_F}=1.3\times 10^{-6}$) and also improves confidence intervals for $K_s$ and $K'_s$. This revealing that the amount of data is related but is not determining to reduce sloppiness or improve identifiability as it had been discussed in previous works (\cite{chandra-transtrum-sethna:2011}). Results are more sensitive to the location of sampling times.

To analyze the effect of the amount of experimental noise we proceeded as described in section 2.5 for the case with constant glucose feeding and $30$ non equidistant sampling times (Case 1c). Figure 3a presents the distribution of $\cal C_F$ values achieved; Figure 3b presents the distribution of the mean of the confidence intervals. Results reveal that both sloppiness and confidence intervals depend on the amount of experimental noise. In general, the larger the variance, the greater the sloppiness and the confidence intervals. 

\begin{center}
\epsffile{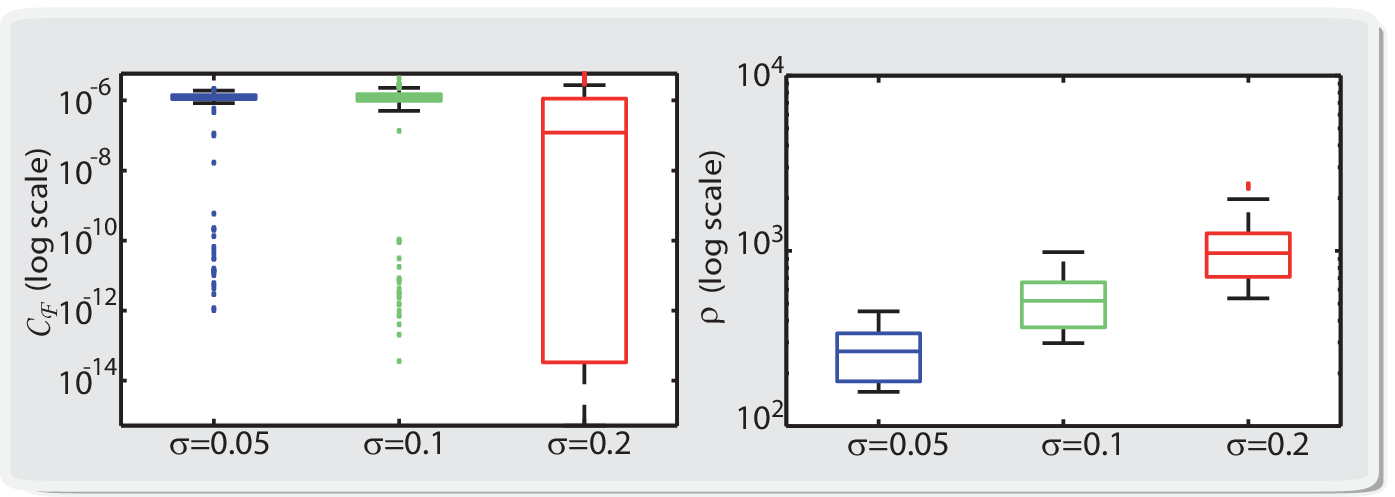} \\
\end{center}
{\bf Figure 3. Role of the experimental noise in the sloppiness and practical identifiability for the ethanol production example under the scenario Case 1c} {\small a) Distribution of the values of ${\cal C}_F$ (log scale) out of $200$ runs for each of the three different experimental noise scenarios. b) Distribution of the mean value of the confidence intervals $\rho$ (log scale) as a measure of the practical identifiability. }

\subsubsection*{Example 5}

For the Example 5, related to a linear biochemical pathway, we considered the stimulus is set to $u=1$ and the experiment duration is $t_f=20$. Since all $16$ parameters are to be estimated a reasonable number of data was selected to be $56$. In case 1a), only first and last components are measured, $28$ samples each; in case 1b) half the components are measured, $8$ samples each; from the more than $3000$ different combinations we selected to measure ($x_1$, $x_2$, $x_5$, $x_7$, $x_{10}$, $x_{13}$ and $x_{14}$) and in case 1c) all components in the network are measured, $4$ samples each. In all cases, sampling times are selected to be equidistantly distributed.

Figure 4 presents an overview of the results. Figure 4a accounts for the sloppiness of the system. Results show how the system is sloppy in all scenarios with ${\cal C_F}<1\times 10^{-5}$. However the distribution of eigenvalues and the values of ${\cal C_F}$ change with the number of observables. The possibility of measuring all the components in the pathway significantly reduced the sloppiness of the system and improved the confidence on the parameter values (Figure 4b). Note that this is a characteristic of linear pathways. In this respect, Figure 4c shows the sensitivity of the different observables with respect to the different parameters illustrating how the sensitivity matrix is almost triangular, as expected. Figure 4c also reveals that it would have been necessary to measure $x_3$, $x_4$ or $x_6$, $x_9$ and $x_{12}$ to improve the identifiability of $p_3$, $p_4$, $p_9$ and $p_12$. The addition of observables improves confidence (Figure 4b) and tends to decorrelate parameters (Figures 4d).

\begin{center}
\epsffile{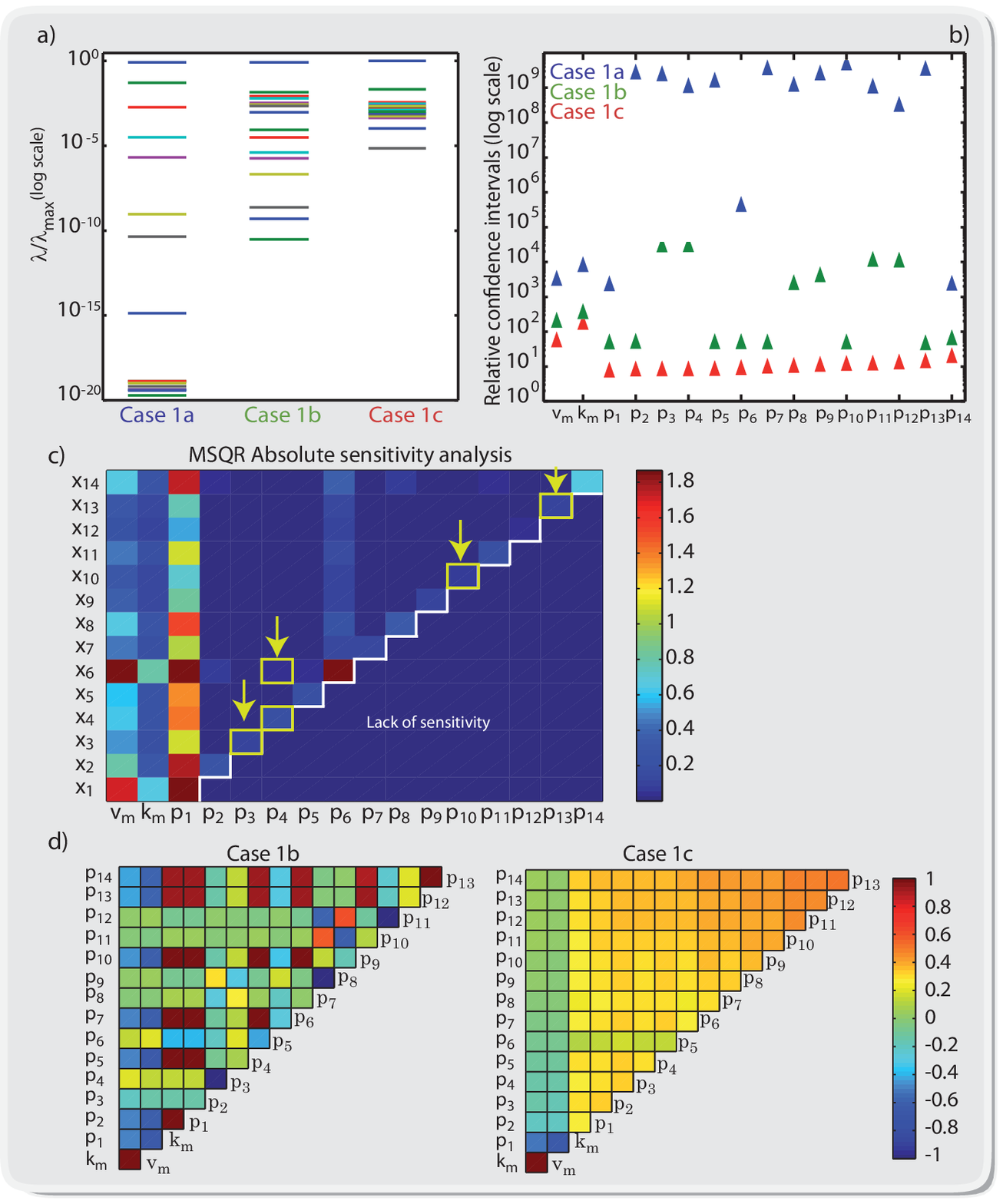} \\
\end{center}
{\bf Figure 4. Sloppiness vs practical identifiability for the linear pathway example.} {\small a) Normalized eigenvalue spectra as a measure of the sloppiness for the different scenarios. The eigenvalues span around 20 orders of magnitude, for the case where only the first and the last component of the network are measured and around 6 for the case all components are measured, indicating sloppiness in all scenarios. b) Normalized confidence intervals for the different model parameters for the given different scenarios. Results show how confidence intervals improve with the addition of observables. c) Sensitivity matrix to illustrate how the different observables depend on the parameters. All observables are sensitive to  $p_1$, thus its value is computed with the highest confidence. $x_6$, $x_1$, $x_14$ and $x_2$ are the most informative observables for the purpose of parameter estimation. d) Correlation matrix for the Case 1b and Case 1c, showing how decorrelation increases with the addition of observables.}

Figure 5 presents the effect of the amount of experimental noise for the scenario 1c. As for the previous example, the value of ${\cal C}_F$ is influenced by the amount of experimental noise and all cases would be regarded as sloppy. The mean value of the confidence intervals also increases with the experimental error, as expected.

\begin{center}
\epsffile{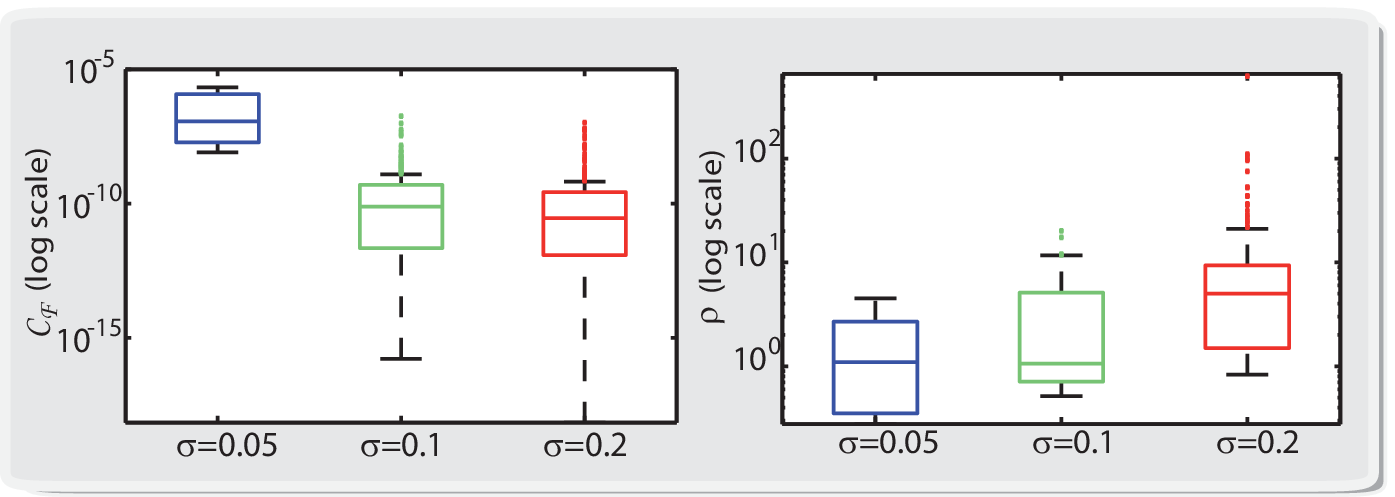} \\
\end{center}
{\bf Figure 5. Role of the experimental noise in the sloppiness and practical identifiability for the linear pathway example} {\small a) Distribution of the values of $log({\cal C}_F)$ out of $200$ runs for each of the three different experimental noise scenarios. Values expand several orders of magnitude. In all cases model is sloppy. b) Distribution of the mean value of the confidence intervals $\rho$ as a measure of the practical identifiability. Values expand several orders of magnitude. Confidence intervals are larger for the cases with larger variance in the experimental error. }

\subsubsection*{Example 6}
For the Example 6, related to a six-gene regulatory network, time courses of all mRNA concentrations may be obtained by microarray experiments and protein concentrations can be measured by means of fluorescence microscope experiments. In a first scenario we assumed that all mRNA and protein concentrations were measured for the wild-type case. The experiment was considered to last $20$ time units and $21$ equidistant times were assumed ($252$ total data). Experimental error is assumed to have $\sigma=0.20$. Results of a sensitivity analysis under those conditions anticipates problems to estimate several parameters, specially binding affinities and Hill coefficients (Figure 6a). 

In fact, the solution of the parameter estimation problem, for $200$ different realizations of the experimental data, reveals large parameter value distributions (Figure 6b). This reveals poor practical identifiability. In addition, ${\cal C}_F$ values are within the range $1\times 10^{-20}-1\times  10^{-12}$ indicating that the model is sloppy independently of the realization of the data.

\begin{center}
\epsffile{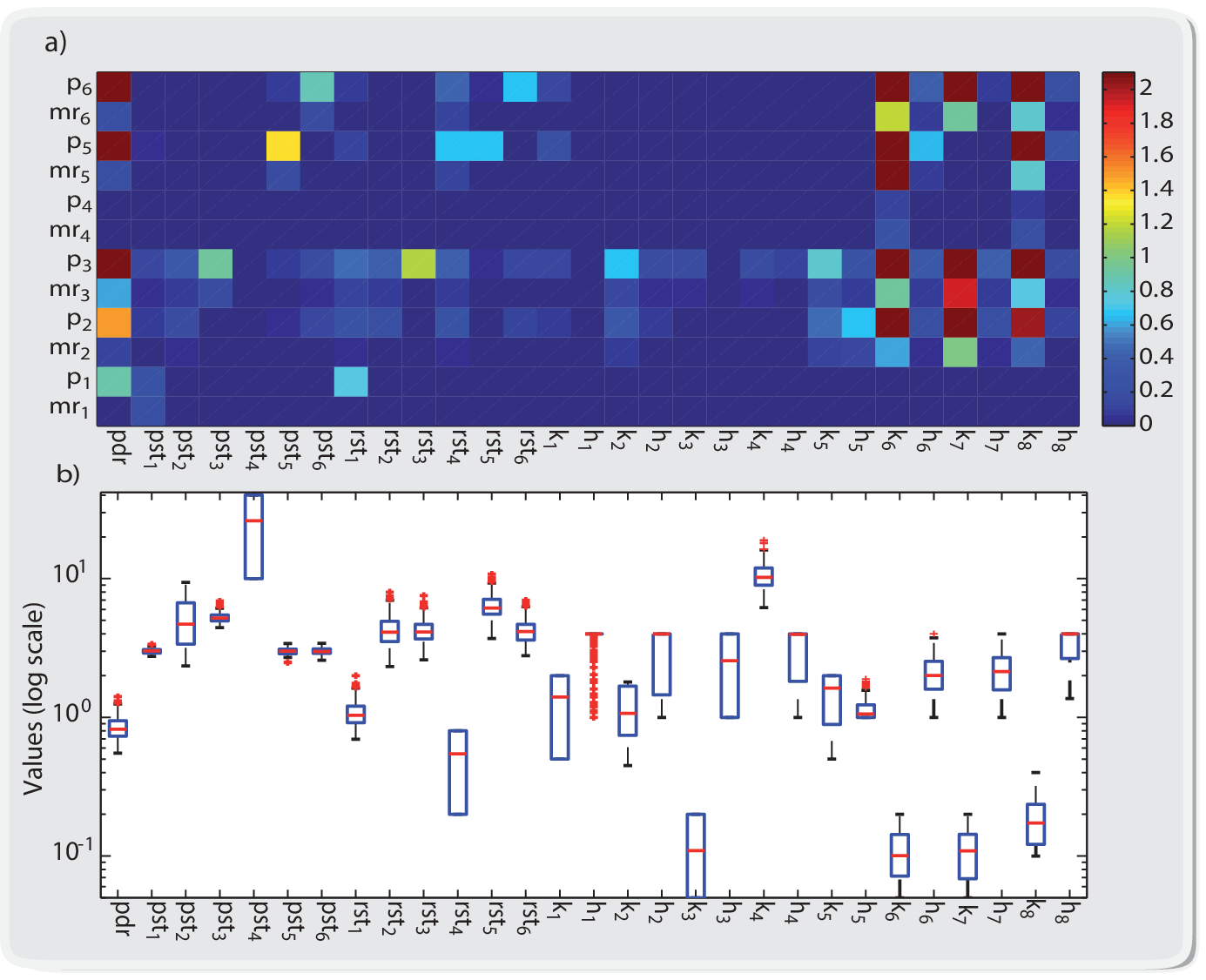} \\
\end{center}
{\bf Figure 6. Role of the experimental noise in the sloppiness and practical identifiability for the 6-gene regulatory network} {\small a) Sensitivity matrix showing how some observables are almost not sensitive to parameter changes under wild type conditions ($mr_1$, $mr_4$ or $p_4$) and how some parameters do almost not influence model outputs ($pst_4$, $h_1$, $k_3$, $h_3$, $k_4$ or $h_4$). b) Distribution of optimal parameter values achieved out of $200$ realizations of experimental data. Most of the parameters cannot be estimated with precision.}

\subsection{The role of optimal experimental design and sloppiness}

Previous examples have already shown the importance of the experimental scheme definition in the sloppiness and the practical identifiability of the model at hand. A right choice of the observables (when possible), the location of the sampling times or the stimulation conditions, can transform a model from identifiable to non-identifiable and, in some cases, from sloppy to non-sloppy. 

However the selection of the right experimental conditions may be contra intuitive. In this sense, optimal experimental design techniques enable the possibility of automatically designing the right experiment for a specific objective. The objective can be related to identifiability, i.e. minimizing the parameters confidence intervals or correlation among parameters, or, for example, to minimize sloppiness.  

To analyze the impact of the experiment design we have considered Examples 5 and 6. The associated D-optimum, E-optimum and S-optimum experimental design problems were solved under different experimental constraints and results were compared. 

\subsubsection*{Example 5}

We considered that we had already performed experiment Case 1b. Next experiments were designed to complement the information provided by that first experiment. For the optimal experimental design we considered the following degrees of freedom: the observed quantities with a maximum of $7$; the stimulation conditions ($0\leq u(t)\leq 2$) and the experiment durations ($15 \leq t_f \leq 25$). Sampling times are assumed to be equidistant. The total amount of data is $n^{e}\times 56$. 

The addition of one single experiment was able to reduce the sloppiness in up to $5$ orders of magnitude, with respect to the Case 1b scenario, in all experimental designs. A further reduction could be achieved by adding an extra designed experiment. However the addition of further experiments did not lead us to a non-sloppy scenario. In any case, despite being sloppy for all experimental designs, confidence intervals for the parameters were substantially reduced for all parameters (see Figure 7) to the point that all of them are considered practically identifiable. Note for example that for the $E-opt$ experimental design, the mean confidence interval is $\rho=4.7\%$ whereas the maximum is $max(\rho_i)=10.5\%$.

\begin{center}
\epsffile{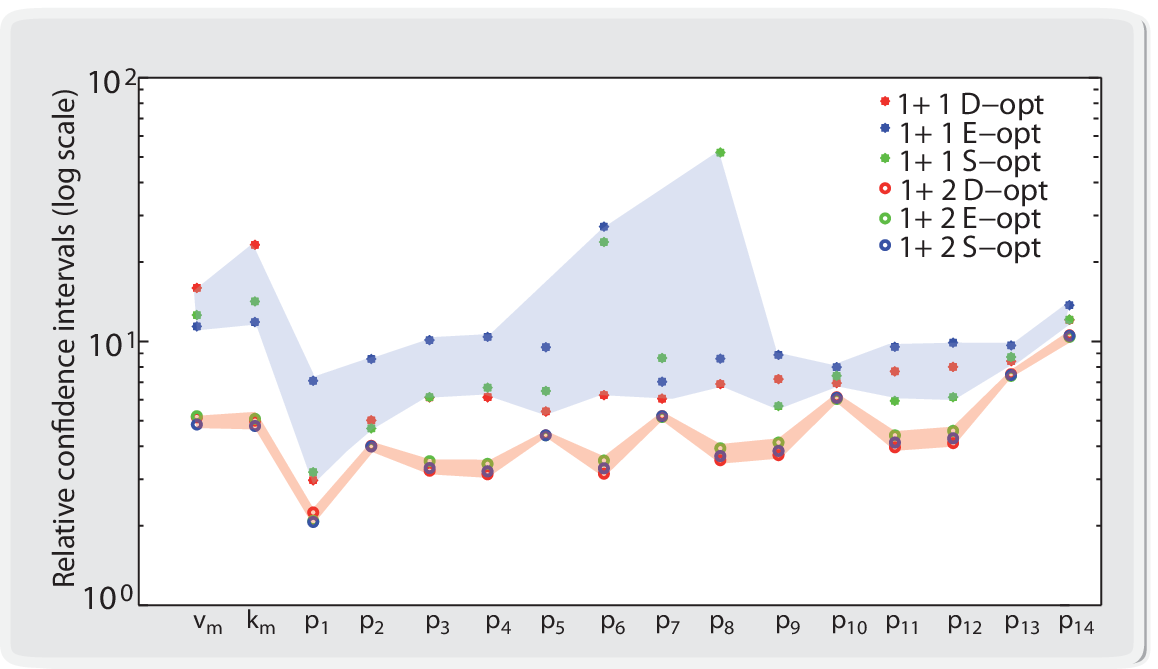} \\
\end{center}
{\bf Figure 7. Evolution of the relative confidence intervals with optimal experimental design for the linear pathway.} {\small Stars present results obtained by Case b experiment plus one optimally designed experiment. The use of different criteria for optimal experimental design results in substantial differences for the confidence intervals (wide blue shaded area). Circles present results achived with Case b plus two optimally designed experiments. All criteria converge to practically the same result (thin orange shaded area), due to the imposed experimental constraints.}

The optimal experimental designs for the cases with two designed experiments are shown in Figure 8. D-opt and E-opt criteria resulted in pretty similar experiments. The optimum for both criteria is to measure $x_3, x_4, x_6, x_8, x_9, x_{11}, x_{12}$ as already expected from the sensitivity analysis (Figure 4c). Feeding profiles seem to test several scenarios to complement the information from experiment Case 1b ($u=1$). Two time varying profiles appear were the feedings are either clearly over or clearly bellow $1$. S-opt design resulted in the measurement of $x_1, x_2, x_3, x_4, x_9, x_{11}, x_{12}$ in the first experiment and $x_3, x_4, x_6, x_8, x_9, x_{11}, x_{12}$ in the second. The feeding is maintained close to zero in both cases, even though profiles are selected in such a way that resuls in different dynamic behaviours of the system. 

\begin{center}
\epsffile{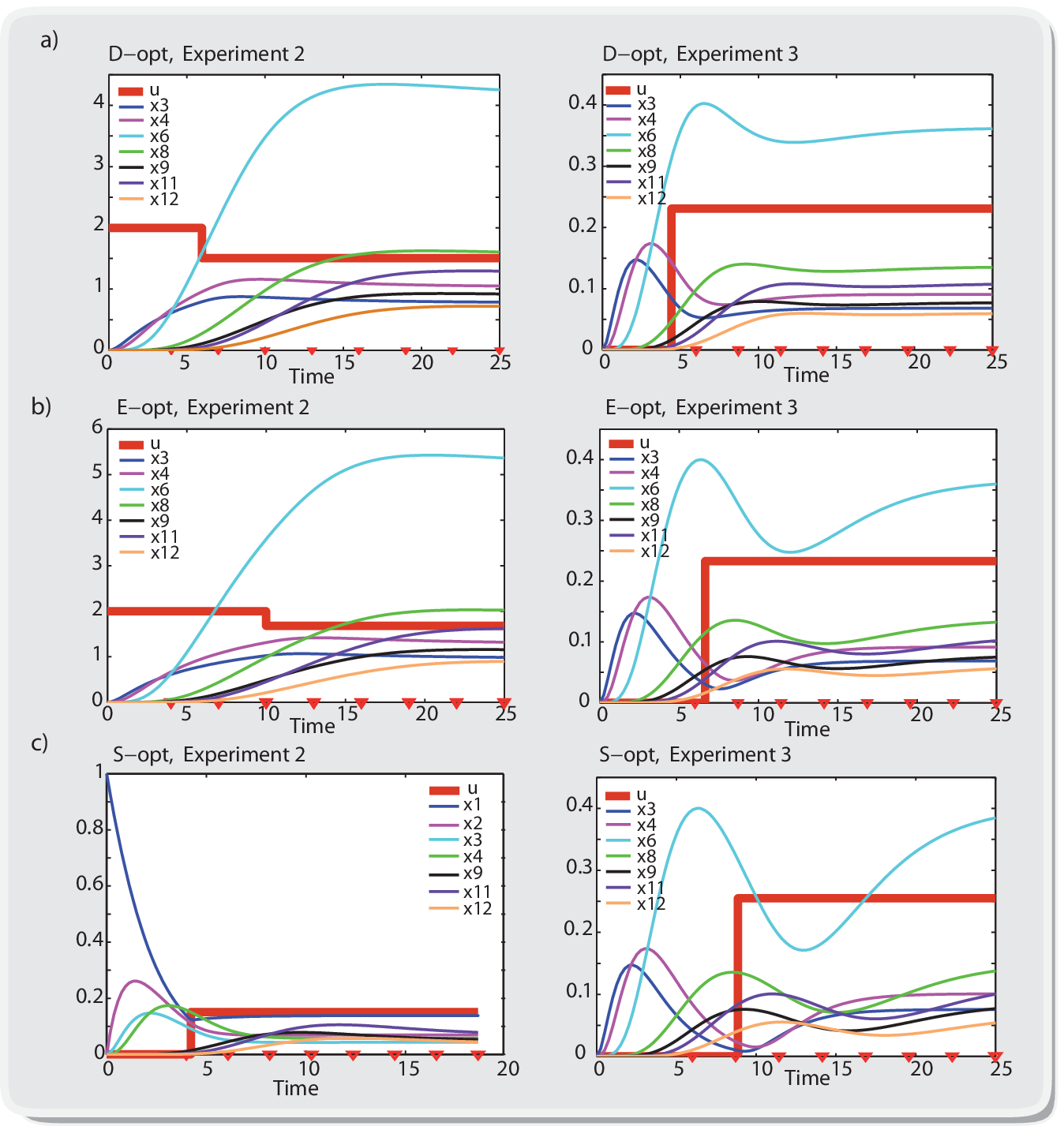} \\
\end{center}
{\bf Figure 8. Optimally designed experiments for the linear pathway example.} {\small Designed experiments for the different criteria: a) D-opt, b) E-opt and c) S-opt. Lines represent stimulation conditions and measured states in each case, triangles in the x-axis represent sampling times.}

\subsubsection*{Example 6}

For the optimal experimental design we assumed (as for the DREAM6 parameter estimation challenge) that the following perturbations can be implemented:

\begin{itemize}
\item GD: Gene deletion to eliminate both mRNA and protein production for a specified gene.
\item KD: mRNA knockdown using siRNA to achieve a 5-fold increase in mRNA degradation.
\item RBA: Increase of RBS activity by $100\%$ (i.e. double) to increase translation rate.
\end{itemize}

For every perturbation we can measure either all mRNA time courses, by means of microarray measurements (M), or time courses of protein abundance for a maximum of 2 proteins per experiment, by means of fluorescent protein fusion (P). For mRNA measurements $21\times 6$ data are obtained whereas for protein measurements only $41\times 2$ data are obtained per experiment. In addition a gel-shift assay (GS) could be used to determine the binding affinity (Kd) and Hill coefficient (h) of any one transcription factor.

The cost of the different experiments is also different, taking as a reference an unit cost $C1$, final costs could be computed as follows:
\begin{itemize}
\item GD: $2\times C1$, KD: $1\times C1$ and RBA:  $1\times C1$
\item M:  $2\times C1$, P: $1\times C1$
\item GS: $3\times C1$
\end{itemize}

We considered a constraint on the maximum budget of $20\times C1$ cost. D-optimum, E-optimum and S-optimum designs were computed considering the experimental constraints and the maximum allowed budget but also the possibility of reducing the number of perturbations to be implemented. Optimal designs are shown in Table 1.
\begin{table}[hhh]
\begin{center}
\begin{tabular}{|l|l|c|c|c|}
\hline\hline
 Criterion & Sequence of experiments & ${\cal C_F}$ & $\rho$& max($\rho_i$)\\ \hline\hline
 D-opt     & mRNA 5 knockdown, all mRNA measured & $2.1\times 10^{-11}$  & $2451\%$    & $4.726\times 10^4\%$\\
           & Gen 1 deletion, all mRNA measured & $1.2\times 10^{-7}$  & $50.5\%$    & $125\%$\\ 
           & mRNA 5 knockdown, all protein measured & $7.8\times 10^{-8}$  & $37.7\%$    & $122\%$\\ 
           & Gen 1 deletion, all protein measured & $8.6\times 10^{-8}$  & $32.5\%$    & $104\%$\\ 
           & mRNA 1 knockdown, all mRNA measured & $2.7\times 10^{-7}$  & $26.6\%$    & $80.9\%$\\
           & Increase of RBS 1 activity, all mRNA measured & $3.1\times 10^{-7}$ & $24.3\%$ & $77.6\%$ \\
           & Increase of RBS 1 activity, p2 \& p3 measured & $\bm{3.2\times 10^{-7}}$ & $\bm{23.3\%}$ & $\bm{76.1\%}$\\ \hline \hline
 E-opt     & mRNA 5 knockdown, all protein measured & $4.7\times 10^{-10}$  & $481\%$    & $6987\%$    \\ 
           & Gell shift ($h_1$, $k_1$)    & $1.3\times 10^{-9}$  & $418\%$    & $6897\%$    \\
           & mRNa 5 knockdown, all mRNA measured  & $1.8\times 10^{-8}$ & $111\%$ & $1721\%$ \\
           & Gen 1 deletion, all mRNA measured    & $2.9\times 10^{-7}$ & $29.6\%$  & $108\%$ \\
           & Gell shift ($h_3$, $k_3$)    &  $3.7\times 10^{-7}$        & $23.7\%$ & $64.9\%$ \\
           & Gell shift ($h_4$, $k_4$)    &  $5.6\times 10^{-7}$        & $21.3\%$ & $64.9\%$ \\
           & Gen 1 deletion, p3 \& p5 measured & $\bm{8.3\times 10^{-7}}$    & $\bm{16.9\%}$ & $\bm{57.8\%}$ \\ \hline \hline
 S-opt     & Gen 5 deletion, all protein measured   & $7.3\times 10^{-9}$ & $735\%$ & $1.0\times 10^4\%$  \\ 
           & Gen 1 deletion, all mRNA measured      & $5.6\times 10^{-8}$ & $67.1\%$ & $252\%$ \\
           & Gen 5 deletion, all mRNA measured      & $6.1\times 10^{-7}$ & $50.2\%$ & $215\%$  \\ 
           & Gell shift ($h_1$, $k_1$)   &            $6.1\times 10^{-6}$ & $36.3\%$ & $122\%$ \\
           & Gell shift ($h_6$, $k_6$)   &          $1.3\times 10^{-5}$ & $32.3\%$ & $122\%$\\
           & Gen 1 deletion, p1, p2, p3 \& p5 measured & $\bm{2.0\times 10^{-5}}$ & $\bm{22.8\%}$ & $\bm{80.5\%}$ \\ \hline \hline 
\end{tabular}
\end{center}
{\bf Table 1. Optimal design of experiments for the sis-gene regulatory network.} {First column regards the criterion used for optimal experimental design; second column presents the experiments; third column shows the sloppiness for the given experimental scheme and fourth and fifth columns, the corresponding mean and max relative confidence intervals, respectively.}
\end{table}

Results reveal that it was not possible to obtain a non-sloppy scenario under the imposed constraints. Selected experiments vary significantly depending on the criteria. D-optimum design avoids gell shift assays, since those experiments, reduce the dimension and thus the determinant of $\cal{F}$. The tendency is to use all other types of experiments and to meassure all states, when enough budget is available. E-optimum desing exploits a combination of gell shift assays with mRNA knockdown and gen deletion. S-optimum design combines gen deletion and gell shift assays.    

The best result in the sense of identifiability was achieved by E-optimum design. It should be noted that the slopiness for this desing is two orders of magnitude worse than the one achieved by the S-optimum design. 

\begin{center}
\epsffile{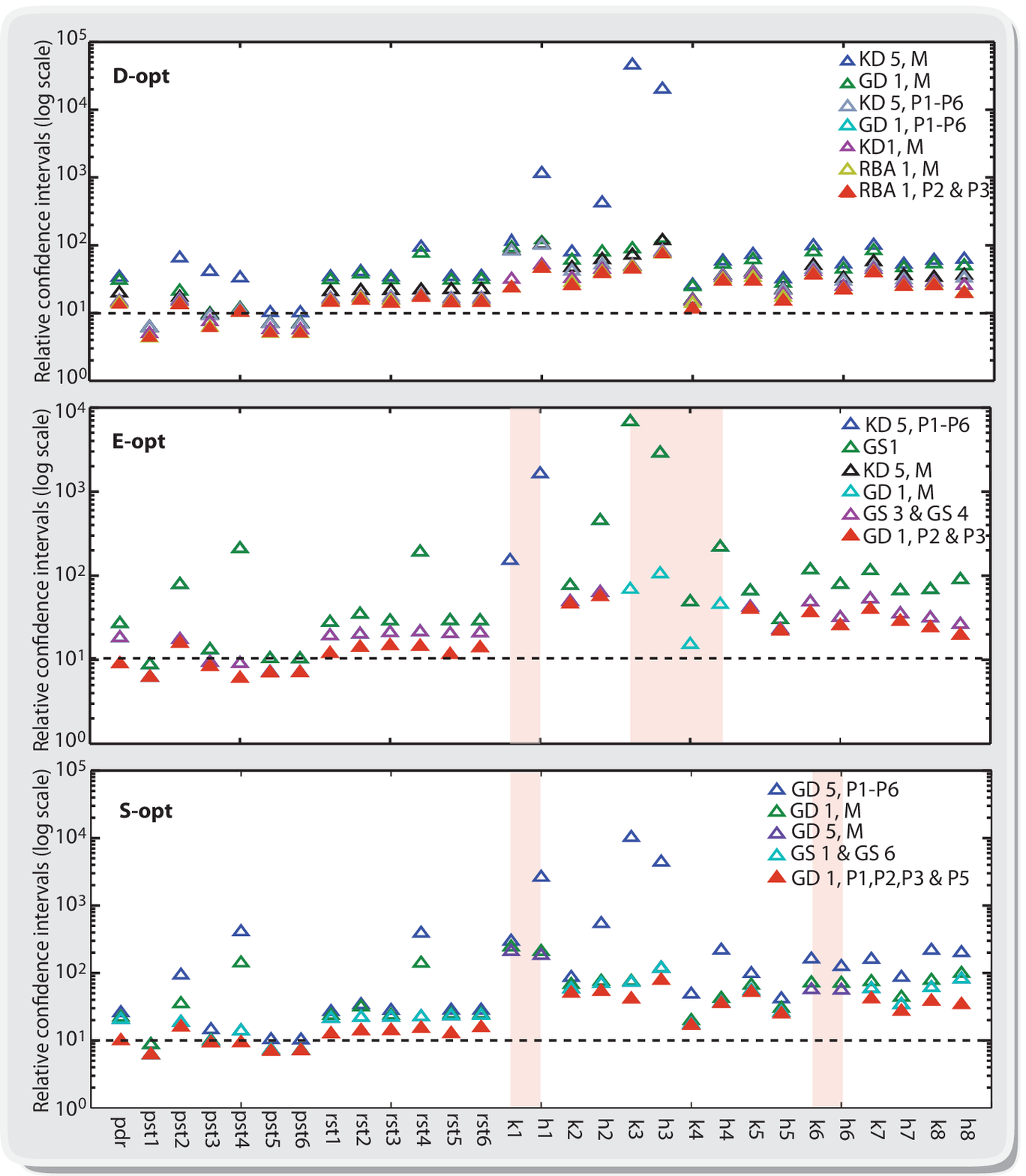} \\
\end{center}
{\bf Figure 9. Evolution of relative confidence intervals in the optimal experimental design for the six-gene regulatory network example.} {\small Figures present the evolution of relative confidence intervals for D-opt, E-opt and S-opt experimental designs. Shaded areas represent parameters that could be measured by means of gell shift experiments, thus being their confidence intervals $0\%$. Red triangles represent the best value achieved. Best overall result corresponds to the E-opt design for which the confidence on the protein degradation rate is bellow the $10\%$; for all promoter strengths and the strengths of the ribosomal binding sites, the relative confidence is around $10\%$ and for binding affinities and Hill coefficients, bellow $60\%$.}

\section{DISCUSSION}

Recent works by Sethna and co-workers \cite{gutenkunst-etal-1:2007,gutenkunst-etal:2007} stated that sloppiness is an inherent property of systems biology models. In this regard the structural identifiability analysis of a collection of examples was used in this work to assess to what extent ``inherent'' could be equivalent to ``structural''. Results for a set of examples with different characteristics (size, type of non-linearity), reflected that, in fact, this is not the case, being all models considered sloppy and structurally identifiable.  In this sense sloppiness is not equivalent to lack of structural identifiability. This means that under ideal noise free experimental data it would be possible to find unique values for the parameters, which contradicts what had been stated in the sloppiness literature cited above

Since the origin of sloppiness is not clearly, or at least not exclusively, related to model structure, it should be rooted to something else. Therefore in the scenario of a sloppy structuraly identifiable model, the question is not whether it is possible to give unique values for the parameters but whether it is possible to compute values for the parameters with confidence, given a set of noisy experimental data. In this concern, several works have commented on the equivalence between sloppiness and poor or lack of practical identifiability (see, for example, \cite{apgar:2010,raue:2011,cirit:2012}). The practical identifiability results will in general depend on i) the experimental conditions (number of experiments, number of observables, number and location of sampling times, initial state of the system and stimulation conditions) and ii) the type and amount of experimental noise. 

Recent works \cite{balsa-alonso-banga:2008,apgar:2010} show how an adequate experimental design may improve the confidence on parameter estimates. In their work \cite{apgar:2010} analyzed the evolution of the sloppiness with the number of designed experiments. Their results suggested that the sloppiness can be reduced significantly with a careful experimental design. Later, in their comment to Agpar et al. work, Chandra et al. \cite{chandra-transtrum-sethna:2011} argue that i) after experimental design the model is still sloppy and ii) that the amount of data required is too high. 

Our results show that sloppiness and confidence intervals vary substantially with experimental noise. The solution of a parameter estimation problem under different realizations of the experimental data results in a distribution of parameter values, thus sloppiness values. Similarly, influence of the selected experimental scheme is also important. Results reveal that the possibility of adding observables tends to reduce sloppiness whereas the amount of data is not as critical as the adequate selection of sampling times. It should be noted, however, that even they may be related, sloppiness and practical identifiability are not equivalent. In fact we show, with several examples, that a given experimental scheme may result in sloppiness whereas the confidence intervals of the parameters are reasonable, and thus we would consider the model practically identifiable.  

Model based optimal experimental design may contribute to reduce sloppiness. We tested the possibility to design experiments to minimise sloppiness and compared results with conventional D-optimum and E-optimum designs. The distribution of relative confidence intervals is different for the different optimization criteria, as expected. And so is sloppiness. In fact, what we observed is that, designs which minimize sloppiness do not end up in the best compromise between mean and maximum expected confidence on the parameters. The additive nature of the Fisher information matrix, implies that the addition of informative data will increase the determinant of $\cal{F}$, thus reducing the mean of the confidence intervals, but it does not imply that the ratio between the minimum and the maximum eigenvalue will be affected. This can be easily seen from a geometric perspective (See the Supplementary info for more details). Let's assume that we perform an experiment that is able to reduce to half all semi axis of the confidence hyper-ellipsoid. It is clear that confidence intervals will be reduced for all parameters, improving identifiability, whereas sloppiness remains the same. Taking this into consideration, it is suggested to use E-optimum or D-optimum designs instead of pursuing the minimum sloppiness design.  

It should be remarked at this point that, experimental constraints and noise may imply the impossibility to transform a sloppy model into a non-sloppy model, but this does not necessarily mean that reliable estimates of the parameters cannot be computed. Since sloppiness is not equivalent to lack or poor practical identifiability, the use of sloppiness as the criterion to decide whether parameters can be uniquely estimated may be misleading. In this sense, performing a practical identifiability analysis, for example, by means of a non-parametric bootstrap approach, would be advisable, before taking a decision on whether the parameters can or cannot be estimated.

\section{CONCLUSIONS}

This work addressed the possible sources and consequences of sloppiness. We considered a collection of examples related to biochemical networks with different characteristics (dimension and non-linearity) under different experimental scenarios, including optimally designed experiments. Structural identifiability analysis of sloppy models indicates that sloppiness is not or, at least not exclusively, related to model structure. Further analyses have led us to conclude that the sloppiness also originates in the selected experimental scheme and the amount of experimental noise. In this regard, we also showed that optimal experimental design may contribute to reduce sloppiness. However, a reduction in the sloppiness of the model  does not necessarily mean an improvement of practical identifiability. In this sense, E-opt designs seem to be more appropriate.    

The idea that a good fit is a guarantee for the validity of the model and thus the validity of the corresponding predictions, although tempting, requires further analyses. Similarly, the idea that sloppy parameters can take any value without modifying model predictions, althoug tempting, does also require further analyses. 

To sum up, our results show that sloppiness is neither equivalent to lack of structural nor to lack or poor practical identifiability. In consequece, the fact that a model is sloppy does not imply that its parameters cannot be estimated with reasonable confidence or cannot be estimated at all. In this sense analysing the sloppiness of a model cannot substitute a structural or a practical identifiability analysis. In summary, identifiability analysis is a more sound approach to assess the quality of parameter estimation studies and the predictive capabilities of the calibrated models.

\section*{Acknowledgments}

This research received financial support from the Spanish Ministerio de Econom\`{\i}a y Competitividad (and
the FEDER) through the project MultiScales (DPI2011-28112-C04-03), and from the CSIC intramural
project BioREDES (PIE-201170E018).


\bibliographystyle{plain}          
\bibliography{pselab}

\newpage

\begin{flushleft}
{\Large
\textbf{Supplementary info: Sloppy models can be identifiable.}
}
\\[0.5cm]
Oana-Teodora Chis, Julio R. Banga, Eva Balsa-Canto$^*$ \\
$^*$E-mail: ebalsa@iim.csic.es
\end{flushleft}

\subsection*{Details on the structural identifiability analysis of examples}

\subsubsection*{Case study 1: The eukaryotic mitotis  \cite{goldbeter:1991}}

The eukaryotic mitosis described by \cite{goldbeter:1991} is a minimal model for the mitotic oscillator, based on the cascade of post-translational modification that modulates the activity of cdc2 kinase during the cell cycle. The dynamics is described by three state variables, 13 parameters, as explained in \cite{goldbeter:1991}. All states can be measured. The equivalent polynomial form of the model results:
\begin{equation}\label{goldbeter}
\begin{array}{ll}
  \frac{dC}{dt}=v_i-v_dXA_1-k_dC,\\
  \frac{dM}{dt}=V_{M1}CA_2(1-M)A_3-V_2MA_4,\\
  \frac{dX}{dt}=V_{M3}M(1-X)A_5-V_4XA_6,\\
  \frac{dA_1}{dt}=-A_1^2(v_i-v_dXA_1-k_dC),\\
  \frac{dA_2}{dt}=-A_2^2(v_i-v_dXA_1-k_dC),\\
  \frac{dA_3}{dt}=A_3^2(V_{M1}CA_2(1-M)A_3-V_2MA_4),\\
  \frac{dA_4}{dt}=-A_4^2(V_{M1}CA_2(1-M)A_3-V_2MA_4),\\
  \frac{dA_5}{dt}=A_5^2(V_{M3}M(1-X)A_5-V_4XA_6),\\
  \frac{dA_6}{dt}=-A_6^2(V_{M3}M(1-X)A_5-V_4XA_6).\\
\end{array}
\end{equation}
The structural identifiability test was done using GenSSI software \cite{chis-banga-balsa-canto:2011}. The corresponding input file is the following:

\begin{center}
\line(1,0){420}
{\small
\begin{verbatim}
                syms x1 x2 x3 x4 x5 x6 x7 x8 x9 x01 x02 x03 
                syms p1 p2 p3 p4 p5 p6 p7 p8 p9 p10 p11 p12 p13
                
                disp('Number of derivatives'); Nder=4
                disp('Number of states'); Neq=9
                disp('Number of model parameters'); Npar=13
                disp('Number of controls'); Noc=0
                disp('Number of observables'); Nobs=9
                   X=[x1 x2 x3 x4 x5 x6 x7 x8 x9];
               
                disp('Equations of the model')
                   A1 = p1-p2*x1*x3*x4-p4*x1;
                   A2 = p13*x1*x5*(1-x2)*x6-p7*x2*x7;
                   A3 = p9*x2*(1-x3)*x8-p11*x3*x9;
                   A4 = -x4^2*(p1-p2*x1*x3*x4-p4*x1);
                   A5 = -x5^2*(p1-p2*x1*x3*x4-p4*x1);
                   A6 =  x6^2*(p13*x1*x5*(1-x2)*x6-p7*x2*x7);
                   A7 = -x7^2*(p13*x1*x5*(1-x2)*x6-p7*x2*x7);
                   A8 = x8^2*(p9*x2*(1-x3)*x8-p11*x3*x9);
                   A9 = -x9^2*(p9*x2*(1-x3)*x8-p11*x3*x9);
                   F=[A1 A2 A3 A4 A5 A6 A7 A8 A9];

                disp('Controls')
                    g1=0;g2=0;g3=0;g4=0;g5=0;g6=0;g7=0;g8=0;g9=0;
                    G=[g1 g2 g3 g4 g5 g6 g7 g8 g9];

                disp('Observables')
                    h1=x1;h2=x2;h3=x3;h4=x4;h5=x5;h6=x6;h7=x7;h8=x8;h9=x9;
                    H=[h1 h2 h3 h4 h5 h6 h7 h8 h9];

                disp('Initial conditions')
                   IC=[x01 x01 x01 1/(p3+x01) 1/(p5+x01) 1/(p6+1-x01)...
                      1/(p8+x01) 1/(p10+1-x01) 1/(p12+x01)];

                disp('Parameters')
                   Par=[p1 p2 p3 p4 p5 p6 p13 p7 p8 p9 p10 p11 p12]; 

                GenSSI_generating_series(F,G,Noc,Neq,Nder,Nobs,H,X,IC,Par,result_folder);
\end{verbatim}
}
\vspace*{-0.35cm}
\line(1,0){420}
\end{center}

Figure 1a represents the identifiability \textit{tableau} for the model in polynomial form using generic initial conditions, and Figure 1b with known initial conditions  $C_{0}=0.01,M_{0}=0.01,X_{0}=0.01,$ $\frac{1}{K_d+0.01},\frac{1}{K_c+0.01},\frac{1}{K_1+1-0.01}, \frac{1}{K_2+0.01}, \frac{1}{K_3+1-0.01}, \frac{1}{K_4+0.01}.$

\begin{center}
\epsffile{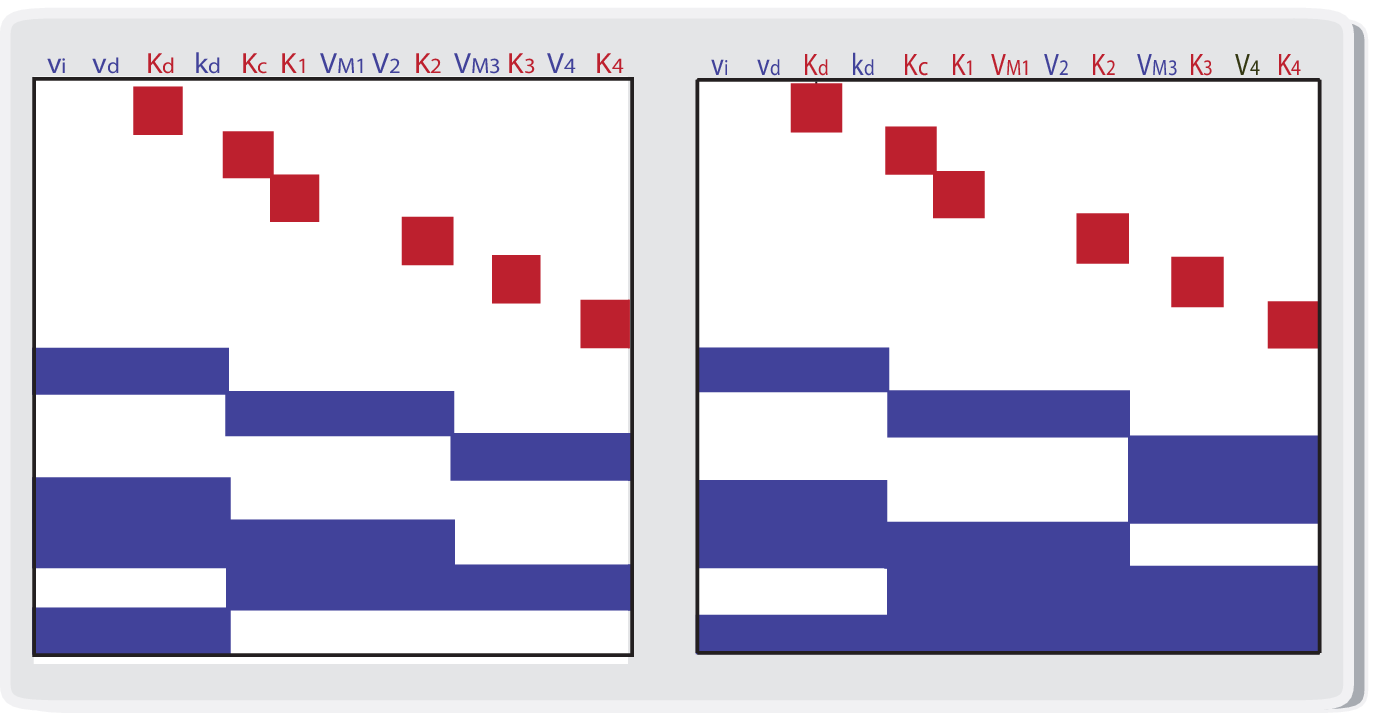} 
\end{center}

\small{
{\bf Figure 1. Identifiability {\em tableaus} for Goldbeter model:} a) generic initial conditions and  b) numerical initial conditions. In red, direct globally identifiable parameters; in blue, at least locally identifiable parameters.}

\newpage

\subsubsection*{Case study 2: The genetic kinase cascade  \cite{kholodenko:2000}}

The the genetic kinase  cascade model is represented by the following system of eight differential kinetic equations (\ref{kholodenko}):

\begin{equation}\label{kholodenko}
\begin{array}{ll}
 \frac{d\,\left[\mathrm{MAPKKK}\right]}{dt} = \frac{\mathrm{V2} \cdot \left[\mathrm{MAPKKK-P}\right]}{\left(\mathrm{K2} + \left[\mathrm{MAPKKK-P}\right]\right)} - \frac{\mathrm{V1} \cdot \left[\mathrm{MAPKKK}\right]}{\left(\left(\frac{\left[\mathrm{MAPK-PP}\right]}{\mathrm{Ki}}\right)^n + 1\right) \cdot \left(\mathrm{K1} + \left[\mathrm{MAPKKK}\right]\right)},\\[3mm]
\frac{d\,\left[\mathrm{MAPKKK-P}\right]}{dt}  = \frac{\mathrm{V1} \cdot \left[\mathrm{MAPKKK}\right]}{\left(\left(\frac{\left[\mathrm{MAPK-PP}\right]}{\mathrm{Ki}}\right)^n + 1\right) \cdot \left(\mathrm{K1} + \left[\mathrm{MAPKKK}\right]\right)} - \frac{\mathrm{V2} \cdot \left[\mathrm{MAPKKK-P}\right]}{\left(\mathrm{K2} + \left[\mathrm{MAPKKK-P}\right]\right)},\\[3mm]
  \frac{d\,\left[\mathrm{MAPKK}\right]}{dt}  =  \frac{\mathrm{V6} \cdot \left[\mathrm{MAPKK-P}\right]}{\left(\mathrm{K6} + \left[\mathrm{MAPKK-P}\right]\right)} - \frac{\mathrm{k3} \cdot \left[\mathrm{MAPKKK-P}\right] \cdot \left[\mathrm{MAPKK}\right]}{\left(\mathrm{K3} + \left[\mathrm{MAPKK}\right]\right)},\\[3mm]
 \frac{d\,\left[\mathrm{MAPKK-P}\right]}{dt}  =  \frac{\mathrm{k3} \cdot \left[\mathrm{MAPKKK-P}\right] \cdot \left[\mathrm{MAPKK}\right]}{\left(\mathrm{K3} + \left[\mathrm{MAPKK}\right]\right)}+  \frac{\mathrm{V5} \cdot \left[\mathrm{MAPKK-PP}\right]}{\left(\mathrm{K5} + \left[\mathrm{MAPKK-PP}\right]\right)} -  \frac{\mathrm{k4} \cdot \left[\mathrm{MAPKKK-P}\right] \cdot \left[\mathrm{MAPKK-P}\right]}{\left(\mathrm{K4} + \left[\mathrm{MAPKK-P}\right]\right)} -  \frac{\mathrm{V6} \cdot \left[\mathrm{MAPKK-P}\right]}{\left(\mathrm{K6} + \left[\mathrm{MAPKK-P}\right]\right)} ,\\[3mm]
\frac{d\,\left[\mathrm{MAPKK-PP}\right]}{dt}  = \frac{\mathrm{k4} \cdot \left[\mathrm{MAPKKK-P}\right] \cdot \left[\mathrm{MAPKK-P}\right]}{\left(\mathrm{K4} + \left[\mathrm{MAPKK-P}\right]\right)} -  \frac{\mathrm{V5} \cdot \left[\mathrm{MAPKK-PP}\right]}{\left(\mathrm{K5} + \left[\mathrm{MAPKK-PP}\right]\right)},\\[3mm]
  \frac{d\,\left[\mathrm{MAPK}\right]}{dt}  = \frac{\mathrm{V10} \cdot \left[\mathrm{MAPK-P}\right]}{\left(\mathrm{K10} + \left[\mathrm{MAPK-P}\right]\right)} - \frac{\mathrm{k7} \cdot \left[\mathrm{MAPKK-PP}\right] \cdot \left[\mathrm{MAPK}\right]}{\left(\mathrm{K7} + \left[\mathrm{MAPK}\right]\right)},\\[3mm]
  \frac{d\,\left[\mathrm{MAPK-P}\right]}{dt}  = \frac{\mathrm{k7} \cdot \left[\mathrm{MAPKK-PP}\right] \cdot \left[\mathrm{MAPK}\right]}{\left(\mathrm{K7} + \left[\mathrm{MAPK}\right]\right)} + \frac{\mathrm{V9} \cdot \left[\mathrm{MAPK-PP}\right]}{\left(\mathrm{K9} + \left[\mathrm{MAPK-PP}\right]\right)} - \frac{\mathrm{k8} \cdot \left[\mathrm{MAPKK-PP}\right] \cdot \left[\mathrm{MAPK-P}\right]}{\left(\mathrm{K8} + \left[\mathrm{MAPK-P}\right]\right)}-  \frac{\mathrm{V10} \cdot \left[\mathrm{MAPK-P}\right]}{\left(\mathrm{K10} + \left[\mathrm{MAPK-P}\right]\right)}, \\[3mm]
 \frac{d\,\left[\mathrm{MAPK-PP}\right]}{dt}  = \frac{\mathrm{k8} \cdot \left[\mathrm{MAPKK-PP}\right] \cdot \left[\mathrm{MAPK-P}\right]}{\left(\mathrm{K8} + \left[\mathrm{MAPK-P}\right]\right)} -  \frac{\mathrm{V9} \cdot \left[\mathrm{MAPK-PP}\right]}{\left(\mathrm{K9} + \left[\mathrm{MAPK-PP}\right]\right)}. 
\end{array}
\end{equation}

GenSSI toolbox was used to perform the identifiability test for the equivalent polynomial model. The parameters K1, K2, K6, K3, K5, K4, K10, K7, K9 and K8 are directly globally identifiable and Ki is locally identifiable having a positive and a negative solution. The remaining parameters n, V1, V5, V6, k7, k8, V9, V10, V2, k3, k4 have also unique solution. The corresponding input file is the following:

\begin{center}
\line(1,0){420}
{\small
\begin{verbatim}
        syms x1 x2 x3 x4 x5 x6 x7 x8 x9 x10 x11 x12 x13 x14 x15 x16 x17 x18 x19 x20
        syms x01 x02 x03 x04 x05 x06 x07 x08
        syms n V1 Ki K1 V2 K2 k3 K3 k4 K4 V5 K5 V6 K6 k7 K7 k8 K8 V9 K9 V10 K10 
    
        disp('Number of derivativs'); 
        Nder=2
        disp('Number of states'); 
        Neq=20
        disp('Number of model parameters'); 
        Npar=22
        disp('Number of controls'); 
        Noc=0
        disp('Number of observables'); 
        Nobs=20
           X=[x1 x2 x3 x4 x5 x6 x7 x8 x9 x10 x11 x12 x13 x14 x15 x16 x17 x18 x19 x20];
        disp('Equations of the model')
           A1 = V2*x2*x12-V1*x1*x10*x11;
           A2 = V1*x1*x10*x11-V2*x2*x12;
           A3 = V6*x4*x13-k3*x2*x3*x14;
           A4 = k3*x2*x3*x14+V5*x5*x15-k4*x2*x4*x16-V6*x4*x13;
           A5 = k4*x2*x4*x16-V5*x5*x15;
           A6 = V10*x7*x17-k7*x5*x6*x18;
           A7 = k7*x5*x6*x18+V9*x8*x19-k8*x5*x7*x20-V10*x7*x17;
           A8 = k8*x5*x7*x20-V9*x8*x19;
           A9 = n*x9*(k8*x5*x7*x20-V9*x8*x19)/x8;
           A10 = -x10^2*(n*x9*(k8*x5*x7*x20-V9*x8*x19)/x8);
           A11 = -x11^2*(V6*x4*x13-k3*x2*x3*x14);
           A12 = -x12^2*(V1*x1*x10*x11-V2*x2*x12);
           A13 = -x13^2*(k3*x2*x3*x14+ V5*x5*x15-k4*x2*x4*x16-V6*x4*x13);
           A14 = -x14^2*(V6*x4*x13-k3*x2*x3*x14);
           A15 = -x15^2*(k4*x2*x4*x16-V5*x5*x15);
           A16 = -x16^2*(k3*x2*x3*x14+V5*x5*x15-k4*x2*x4*x16-V6*x4*x13);
           A17 = -x17^2*(k7*x5*x6*x18+V9*x8*x19-k8*x5*x7*x20-V10*x7*x17);
           A18 = -x18^2*(V10*x7*x17-k7*x5*x6*x18);
           A19 = -x19^2*(k8*x5*x7*x20-V9*x8*x19);
           A20 = -x20^2*(k7*x5*x6*x18+V9*x8*x19-k8*x5*x7*x20-V10*x7*x17);
           F=[A1 A2 A3 A4 A5 A6 A7 A8 A9 A10 A11 A12 A13 A14 A15 A16 A17 A18 A19 A20];

       disp('Controls')
           g1=0;g2=0;g3=0;g4=0;g5=0;g6=0;g7=0;g8=0;g9=0;g10=0;g11=0;                  
           g12=0;g13=0;g14=0;g15=0;g16=0;g17=0;g18=0;g19=0;g20=0;                 
           G=[g1 g2 g3 g4 g5 g6 g7 g8 g9 g10 g11 g12 g13 g14 g15 g16 g17 g18 g19 g20];

       disp('Observables')
           h1=x1;h2=x2;h3=x3;h4=x4;h5=x5;h6=x6;h7=x7;h8=x8;h9=x9;h10=x10;h11=x11;
           h12=x12;h13=x13;h14=x14;h15=x15;h16=x16;h17=x17;h18=x18;h19=x19;h20=x20;
           H=[h1 h2 h3 h4 h5 h6 h7 h8 h9 h10 h11 h12 h13 h14 h15 h16 h17 h18 h19 h20];

       disp('Initial conditions')
           IC=[x01 x02 x03 x04 x05 x06 x07 x08 (x08/Ki)^2 1/(1+(x08/Ki)^2) ...
              1/(K1+x01) 1/(K2+x02) 1/(K6+x04) 1/(K3+x03) 1/(K5+x05) 1/(K4+x04) ...
              1/(K10+x07) 1/(K7+x06) 1/(K9+x08) 1/(K8+x07)];

       disp('Parameters')
           Par=[n, V1, Ki, K1, V2, K2, k3, K3, k4, K4, ...
                V5, K5, V6, K6, k7, K7, k8, K8, V9, K9, V10, K10]; 

       GenSSI_generating_series(F,G,Noc,Neq,Nder,Nobs,H,X,IC,Par,result_folder);
\end{verbatim}
}
\vspace*{-0.35cm}
\line(1,0){420}
\end{center}

\newpage

\subsubsection*{Case study 3: the growth-factor-signalling network in PC12 cells \cite{brown-hill-calero-myers-lee-sethna-cerione:2004}}

This model represents the actions of neuronal growth factor (NGF) and mitogenic epidermal growth
factor (EGF) in rat pheochromocytoma (PC12) cells. These growth factors stimulate extracellular regulated kinase (Erk) phosphorylation using intermediate signalling proteins, with distinct dynamical profiles. The whole process is represented by the following system of differential equations, having 28 state variables and 48 parameters.

\begin{equation}\label{pc21}
\begin{array}{ll}
  \frac{d\,\left[\mathrm{EGF}\right]}{dt} &=  \mathrm{cell} \cdot \mathrm{kruEGF} \cdot \left[\mathrm{boundEGFR}\right] - \mathrm{cell} \cdot \mathrm{krbEGF} \cdot \left[\mathrm{EGF}\right] \cdot \left[\mathrm{freeEGFR}\right] ,\nonumber\\[3mm]
  \frac{d\,\left[\mathrm{NGF}\right]}{dt}  &=  \mathrm{kruNGF} \cdot \left[\mathrm{boundNGFR}\right] \cdot \mathrm{cell} - \mathrm{krbNGF} \cdot \left[\mathrm{NGF}\right] \cdot \left[\mathrm{freeNGFR}\right] \cdot \mathrm{cell},\nonumber\\[3mm]
  \frac{d\,\left[\mathrm{freeEGFR}\right]}{dt}  &=  \mathrm{cell} \cdot \mathrm{kruEGF} \cdot \left[\mathrm{boundEGFR}\right] - \mathrm{cell} \cdot \mathrm{krbEGF} \cdot \left[\mathrm{EGF}\right] \cdot \left[\mathrm{freeEGFR}\right],\nonumber \\[3mm]
  \frac{d\,\left[\mathrm{boundEGFR}\right]}{dt}  &=  \mathrm{cell} \cdot \mathrm{krbEGF} \cdot \left[\mathrm{EGF}\right] \cdot \left[\mathrm{freeEGFR}\right]- \mathrm{cell} \cdot \mathrm{kruEGF} \cdot \left[\mathrm{boundEGFR}\right],\nonumber \\[3mm] 
  \frac{d\,\left[\mathrm{freeNGFR}\right]}{dt}  &=  \mathrm{kruNGF} \cdot \left[\mathrm{boundNGFR}\right] \cdot \mathrm{cell}- \mathrm{krbNGF} \cdot \left[\mathrm{NGF}\right] \cdot \left[\mathrm{freeNGFR}\right] \cdot \mathrm{cell} \nonumber \\[3mm]
  \frac{d\,\left[\mathrm{boundNGFR}\right]}{dt}  &= \mathrm{krbNGF} \cdot \left[\mathrm{NGF}\right] \cdot \left[\mathrm{freeNGFR}\right] \cdot \mathrm{cell} - \mathrm{kruNGF} \cdot \left[\mathrm{boundNGFR}\right] \cdot \mathrm{cell},\nonumber \\[3mm]
  \frac{d\,\left[\mathrm{SosInact}\right]}{dt}  &= \frac{\mathrm{cell} \cdot \mathrm{kdSos} \cdot \left[\mathrm{P90RskAct}\right] \cdot \left[\mathrm{SosAct}\right]}{\left(\left[\mathrm{SosAct}\right] + \mathrm{KmdSos}\right)} - \frac{\mathrm{cell} \cdot \mathrm{kEGF} \cdot \left[\mathrm{boundEGFR}\right] \cdot \left[\mathrm{SosInact}\right]}{\left(\left[\mathrm{SosInact}\right] + \mathrm{KmEGF}\right)}-\frac{\mathrm{cell} \cdot \mathrm{kNGF} \cdot \left[\mathrm{boundNGFR}\right] \cdot \left[\mathrm{SosInact}\right]}{\left(\left[\mathrm{SosInact}\right] + \mathrm{KmNGF}\right)},\nonumber \\[3mm]
  \frac{d\,\left[\mathrm{SosAct}\right]}{dt} &= \frac{\mathrm{cell} \cdot \mathrm{kEGF} \cdot \left[\mathrm{boundEGFReceptor}\right] \cdot \left[\mathrm{SosInact}\right]}{\left(\left[\mathrm{SosInact}\right] + \mathrm{KmEGF}\right)} + \frac{\mathrm{cell} \cdot \mathrm{kNGF} \cdot \left[\mathrm{boundNGFR}\right] \cdot \left[\mathrm{SosInact}\right]}{\left(\left[\mathrm{SosInact}\right] + \mathrm{KmNGF}\right)}-\frac{\mathrm{cell} \cdot \mathrm{kdSos} \cdot \left[\mathrm{P90RskAct}\right] \cdot \left[\mathrm{SosAct}\right]}{\left(\left[\mathrm{SosAct}\right] + \mathrm{KmdSos}\right)},\nonumber\\[3mm]
  \frac{d\,\left[\mathrm{P90RskInact}\right]}{dt}  &= -\frac{\mathrm{cell} \cdot \mathrm{kpP90Rsk} \cdot \left[\mathrm{ErkAct}\right] \cdot \left[\mathrm{P90RskInact}\right]}{\left(\left[\mathrm{P90RskInact}\right] + \mathrm{KmpP90Rsk}\right)}),\nonumber \\[3mm]
  \frac{d\,\left[\mathrm{P90RskAct}\right]}{dt}  &=  \frac{\mathrm{cell} \cdot \mathrm{kpP90Rsk} \cdot \left[\mathrm{ErkAct}\right] \cdot \left[\mathrm{P90RskInact}\right]}{\left(\left[\mathrm{P90RskInact}\right] + \mathrm{KmpP90Rsk}\right)},\nonumber \\[3mm]
 \frac{d\,\left[\mathrm{RasInact}\right]}{dt}  &=  \frac{\mathrm{cell} \cdot \mathrm{kRasGap} \cdot \left[\mathrm{RasGapAct}\right] \cdot \left[\mathrm{RasAct}\right]}{\left(\left[\mathrm{RasAct}\right] + \mathrm{KmRasGap}\right)}- \frac{\mathrm{cell} \cdot \mathrm{kSos} \cdot \left[\mathrm{SosAct}\right] \cdot \left[\mathrm{RasInact}\right]}{\left(\left[\mathrm{RasInact}\right] + \mathrm{KmSos}\right)} \\[3mm]
 \frac{d\,\left[\mathrm{RasAct}\right]}{dt} &=  \frac{\mathrm{cell} \cdot \mathrm{kSos} \cdot \left[\mathrm{SosAct}\right] \cdot \left[\mathrm{RasInact}\right]}{\left(\left[\mathrm{RasInact}\right] + \mathrm{KmSos}\right)} - \frac{\mathrm{cell} \cdot \mathrm{kRasGap} \cdot \left[\mathrm{RasGapAct}\right] \cdot \left[\mathrm{RasAct}\right]}{\left(\left[\mathrm{RasAct}\right] + \mathrm{KmRasGap}\right)},\nonumber \\[3mm]
  \frac{d\,\left[\mathrm{Raf1Inact}\right]}{dt} &= \frac{\mathrm{cell} \cdot \mathrm{kdRaf1} \cdot \left[\mathrm{Raf1PPtase}\right] \cdot \left[\mathrm{Raf1Act}\right]}{\left(\left[\mathrm{Raf1Act}\right] + \mathrm{KmdRaf1}\right)} + \frac{\mathrm{cell} \cdot \mathrm{kdRaf1ByAkt} \cdot \left[\mathrm{AktAct}\right] \cdot \left[\mathrm{Raf1Act}\right]}{\left(\left[\mathrm{Raf1Act}\right] + \mathrm{KmRaf1ByAkt}\right)} - \frac{\mathrm{cell} \cdot \mathrm{kRasToRaf1} \cdot \left[\mathrm{RasAct}\right] \cdot \left[\mathrm{Raf1Inact}\right]}{\left(\left[\mathrm{Raf1Inact}\right] + \mathrm{KmRasToRaf1}\right)},\nonumber \\[3mm]
  \frac{d\,\left[\mathrm{Raf1Act}\right]}{dt}  &= \frac{\mathrm{cell} \cdot \mathrm{kRasToRaf1} \cdot \left[\mathrm{RasAct}\right] \cdot \left[\mathrm{Raf1Inact}\right]}{\left(\left[\mathrm{Raf1Inact}\right] + \mathrm{KmRasToRaf1}\right)} -\frac{\mathrm{cell} \cdot \mathrm{kdRaf1} \cdot \left[\mathrm{Raf1PPtase}\right] \cdot \left[\mathrm{Raf1Act}\right]}{\left(\left[\mathrm{Raf1Act}\right] + \mathrm{KmdRaf1}\right)} - \frac{\mathrm{cell} \cdot \mathrm{kdRaf1ByAkt} \cdot \left[\mathrm{AktAct}\right] \cdot \left[\mathrm{Raf1Act}\right]}{\left(\left[\mathrm{Raf1Act}\right] + \mathrm{KmRaf1ByAkt}\right)} ,\nonumber \\[3mm]
 \frac{d\,\left[\mathrm{BRafInact}\right]}{dt}  &= \frac{\mathrm{cell} \cdot \mathrm{kdBRaf} \cdot \left[\mathrm{Raf1PPtase}\right] \cdot \left[\mathrm{BRafAct}\right]}{\left(\left[\mathrm{BRafAct}\right] + \mathrm{KmdBRaf}\right)} - \frac{\mathrm{cell} \cdot \mathrm{kRap1ToBRaf} \cdot \left[\mathrm{Rap1Act}\right] \cdot \left[\mathrm{BRafInact}\right]}{\left(\left[\mathrm{BRafInact}\right] + \mathrm{KmRap1ToBRaf}\right)},\nonumber\\[3mm]
  \frac{d\,\left[\mathrm{BRafAct}\right]}{dt}  &= \frac{\mathrm{cell} \cdot \mathrm{kRap1ToBRaf} \cdot \left[\mathrm{Rap1Act}\right] \cdot \left[\mathrm{BRafInact}\right]}{\left(\left[\mathrm{BRafInact}\right] + \mathrm{KmRap1ToBRaf}\right)} - \frac{\mathrm{cell} \cdot \mathrm{kdBRaf} \cdot \left[\mathrm{Raf1PPtase}\right] \cdot \left[\mathrm{BRafAct}\right]}{\left(\left[\mathrm{BRafAct}\right] + \mathrm{KmdBRaf}\right)} ,\nonumber\\[3mm]
  \frac{d\,\left[\mathrm{MekInact}\right]}{dt}  &= \frac{\mathrm{cell} \cdot \mathrm{kdMek} \cdot \left[\mathrm{PP2AAct}\right] \cdot \left[\mathrm{MekAct}\right]}{\left(\left[\mathrm{MekAct}\right] + \mathrm{KmdMek}\right)} - \frac{\mathrm{cell} \cdot \mathrm{kpRaf1} \cdot \left[\mathrm{Raf1Act}\right] \cdot \left[\mathrm{MekInact}\right]}{\left(\left[\mathrm{MekInact}\right] + \mathrm{KmpRaf1}\right)} - \frac{\mathrm{cell} \cdot \mathrm{kpBRaf} \cdot \left[\mathrm{BRafAct}\right] \cdot \left[\mathrm{MekInact}\right]}{\left(\left[\mathrm{MekInact}\right] + \mathrm{KmpBRaf}\right)},\nonumber \\[3mm]
  \frac{d\,\left[\mathrm{MekAct}\right]}{dt}  &= \frac{\mathrm{cell} \cdot \mathrm{kpRaf1} \cdot \left[\mathrm{Raf1Act}\right] \cdot \left[\mathrm{MekInact}\right]}{\left(\left[\mathrm{MekInact}\right] + \mathrm{KmpRaf1}\right)} + \frac{\mathrm{cell} \cdot \mathrm{kpBRaf} \cdot \left[\mathrm{BRafAct}\right] \cdot \left[\mathrm{MekInact}\right]}{\left(\left[\mathrm{MekInact}\right] + \mathrm{KmpBRaf}\right)}-\frac{\mathrm{cell} \cdot \mathrm{kdMek} \cdot \left[\mathrm{PP2AAct}\right] \cdot \left[\mathrm{MekAct}\right]}{\left(\left[\mathrm{MekAct}\right] + \mathrm{KmdMek}\right)},\nonumber \\[3mm]
  \frac{d\,\left[\mathrm{ErkInact}\right]}{dt}  &= \frac{\mathrm{cell} \cdot \mathrm{kdErk} \cdot \left[\mathrm{PP2AAct}\right] \cdot \left[\mathrm{ErkAct}\right]}{\left(\left[\mathrm{ErkAct}\right] + \mathrm{KmdErk}\right)}  - \frac{\mathrm{cell} \cdot \mathrm{kpMekCyt} \cdot \left[\mathrm{MekAct}\right] \cdot \left[\mathrm{ErkInact}\right]}{\left(\left[\mathrm{ErkInact}\right] + \mathrm{KmpMekCyt}\right)} ,\nonumber\\[3mm]
  \frac{d\,\left[\mathrm{ErkAct}\right]}{dt}  &= \frac{\mathrm{cell} \cdot \mathrm{kpMekCyt} \cdot \left[\mathrm{MekAct}\right] \cdot \left[\mathrm{ErkInact}\right]}{\left(\left[\mathrm{ErkInact}\right] + \mathrm{KmpMekCyt}\right)} - \frac{\mathrm{cell} \cdot \mathrm{kdErk} \cdot \left[\mathrm{PP2AAct}\right] \cdot \left[\mathrm{ErkAct}\right]}{\left(\left[\mathrm{ErkAct}\right] + \mathrm{KmdErk}\right)}  ,\nonumber\\[3mm]
  \frac{d\,\left[\mathrm{PI3KInact}\right]}{dt}  &= -\frac{\mathrm{cell} \cdot \mathrm{kPI3K} \cdot \left[\mathrm{boundEGFR}\right] \cdot \left[\mathrm{PI3KInact}\right]}{\left(\left[\mathrm{PI3KInact}\right] + \mathrm{KmPI3K}\right)} - \frac{\mathrm{cell} \cdot \mathrm{kPI3KRas} \cdot \left[\mathrm{RasAct}\right] \cdot \left[\mathrm{PI3KInact}\right]}{\left(\left[\mathrm{PI3KInact}\right] + \mathrm{KmPI3KRas}\right)} ,\nonumber \\[3mm]
  \frac{d\,\left[\mathrm{PI3KAct}\right]}{dt}  &= \frac{\mathrm{cell} \cdot \mathrm{kPI3K} \cdot \left[\mathrm{boundEGFR}\right] \cdot \left[\mathrm{PI3KInact}\right]}{\left(\left[\mathrm{PI3KInact}\right] + \mathrm{KmPI3K}\right)} + \frac{\mathrm{cell} \cdot \mathrm{kPI3KRas} \cdot \left[\mathrm{RasAct}\right] \cdot \left[\mathrm{PI3KInact}\right]}{\left(\left[\mathrm{PI3KInact}\right] + \mathrm{KmPI3KRas}\right)} \nonumber \\[3mm]
  \frac{d\,\left[\mathrm{AktInact}\right]}{dt}  &= -\frac{\mathrm{cell} \cdot \mathrm{kAkt} \cdot \left[\mathrm{PI3KAct}\right] \cdot \left[\mathrm{AktInact}\right]}{\left(\left[\mathrm{AktInact}\right] + \mathrm{KmAkt}\right)}  ,\nonumber\\[3mm]
  \frac{d\,\left[\mathrm{AktAct}\right]}{dt}  &= \frac{\mathrm{cell} \cdot \mathrm{kAkt} \cdot \left[\mathrm{PI3KAct}\right] \cdot \left[\mathrm{AktInact}\right]}{\left(\left[\mathrm{AktInact}\right] + \mathrm{KmAkt}\right)} ,\nonumber \\[3mm]
  \frac{d\,\left[\mathrm{C3GInact}\right]}{dt}  &= -\frac{\mathrm{cell} \cdot \mathrm{kC3GNGF} \cdot \left[\mathrm{boundNGFR}\right] \cdot \left[\mathrm{C3GInact}\right]}{\left(\left[\mathrm{C3GInact}\right] + \mathrm{KmC3GNGF}\right)}  ,\nonumber\\[3mm]
  \frac{d\,\left[\mathrm{C3GAct}\right]}{dt}  &= \frac{\mathrm{cell} \cdot \mathrm{kC3GNGF} \cdot \left[\mathrm{boundNGFR}\right] \cdot \left[\mathrm{C3GInact}\right]}{\left(\left[\mathrm{C3GInact}\right] + \mathrm{KmC3GNGF}\right)} ,\nonumber\\[3mm]
  \frac{d\,\left[\mathrm{Rap1Inact}\right]}{dt}  &= \frac{\mathrm{cell} \cdot \mathrm{kRapGap} \cdot \left[\mathrm{RapGapAct}\right] \cdot \left[\mathrm{Rap1Act}\right]}{\left(\left[\mathrm{Rap1Act}\right] + \mathrm{KmRapGap}\right)} - \frac{\mathrm{cell} \cdot \mathrm{kC3G} \cdot \left[\mathrm{C3GAct}\right] \cdot \left[\mathrm{Rap1Inact}\right]}{\left(\left[\mathrm{Rap1Inact}\right] + \mathrm{KmC3G}\right)}  ,\nonumber\\[3mm]
 \frac{d\,\left[\mathrm{Rap1Act}\right]}{dt}  &= \frac{\mathrm{cell} \cdot \mathrm{kC3G} \cdot \left[\mathrm{C3GAct}\right] \cdot \left[\mathrm{Rap1Inact}\right]}{\left(\left[\mathrm{Rap1Inact}\right] + \mathrm{KmC3G}\right)} -\frac{\mathrm{cell} \cdot \mathrm{kRapGap} \cdot \left[\mathrm{RapGapAct}\right] \cdot \left[\mathrm{Rap1Act}\right]}{\left(\left[\mathrm{Rap1Act}\right] + \mathrm{KmRapGap}\right)}  .\nonumber
\end{array}
\end{equation}

The identifiability test was done for the equivalent polynomial model. After 10 derivatives a unique solution for all the parameters was found, thus the Brown model is structurally globally identifiable. In this case the computations were done using generating series approach as implemented in MATHEMATICA.

\subsubsection*{Case study 4: Fed-batch reactor for ethanol production \cite{hong:86}}

Structural identifiability analysis is done for the model (9) described in the main document. The input GenSSI file is given above: 

\begin{center}
\line(1,0){420}
{\small
\begin{verbatim}
                syms x S P V mu0 KP KS KPp KSp nu0 x01 x02 x03 x04 
                disp('Number of derivatives'); Nder=2
                disp('Number of states'); Neq=4
                disp('Number of model parameters'); Npar=6
                disp('Number of controls'); Noc=1
                disp('Number of observables'); Nobs=4
                   X=[x S P V];
                disp('Equations of the model')
                   A1 = x*mu0*S/((KS+S)*(1+P/KP));        
                   A2 = -nu0*x*mu0*S/((KS+S)*(1+P/KP));
                   A3 = x*S/((KSp+S)*(1+P/KPp));
                   A4 =0;
                   F=[A1 A2 A3 A4];
                disp('Controls')
                    g1=-x/V;g2=(150-S)/V;g3=-P/V;g4=1;
                    G=[g1 g2 g3 g4];
                disp('Observables')
                    h1=x;h2=s;h3=p;h4=v;
                    H=[h1 h2 h3 h4];
                disp('Initial conditions')
                   IC=[x01 x02 x03 x04];
                disp('Parameters')
                   Par=[mu0, KP, KS, KPp, KSp, nu0]; 
                GenSSI_generating_series(F,G,Noc,Neq,Nder,Nobs,H,X,IC,Par,result_folder);

\end{verbatim}
}
\vspace*{-0.35cm}
\line(1,0){420}
\end{center}         

The model is locally structurally identifiable, the parameters $\mu_0$, $K_S$, $K_S^{'}$ and $\nu_0$ are globally identifiable and $K_P$ and $K_P^{'}$ are locally identifiable. These results are obtained considering generic initial conditions. The identifiability \textit{tableau} is represented in Figure 2b. If the initial conditions are [$x$=1, $S$=150, $P$=0, $V$=10] GenSSI needs 3 derivatives and in this case the parameters  $K_S^{'}$, $K_P$ and $\nu_0$ are globally identifiable and  $\mu_0$, $K_S$ and $K_P^{'}$ are locally structurally identifiable. The corresponding identifiability \textit{tableau} is given in Figure 2a.

\begin{center}
\epsffile{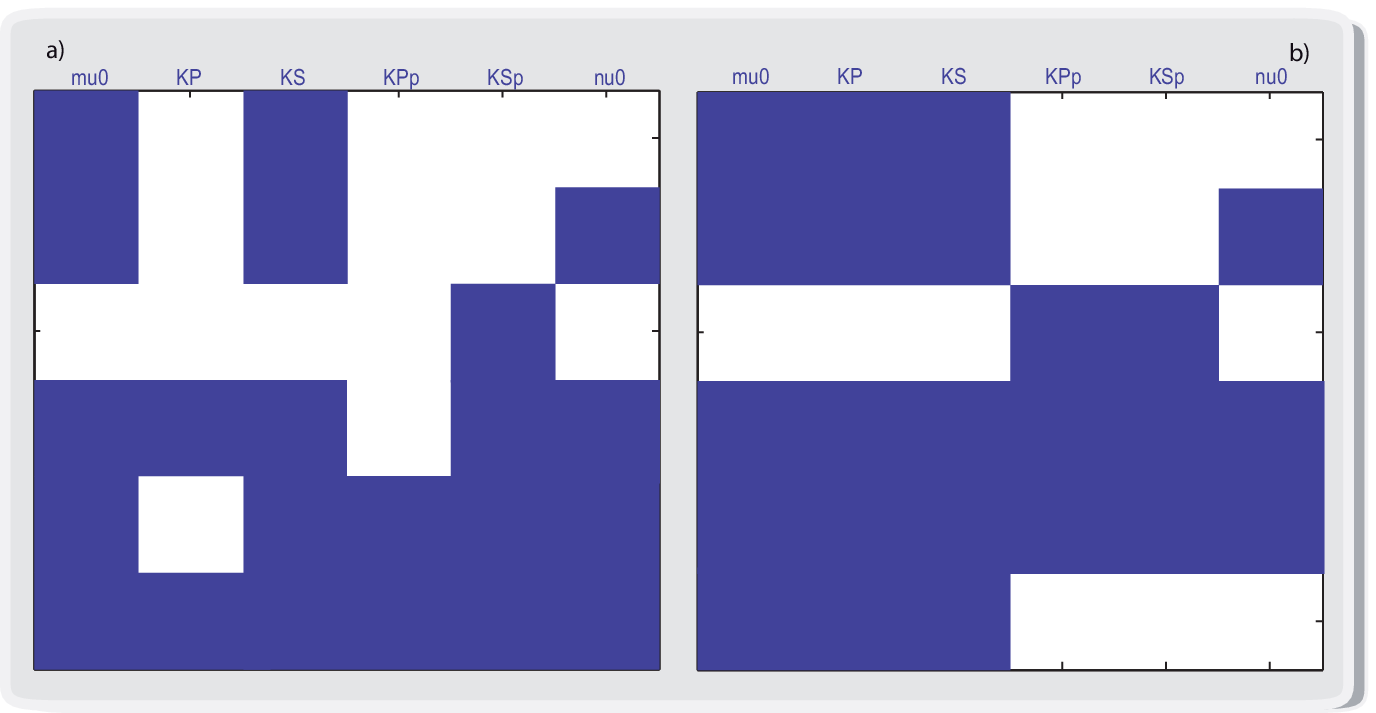} 
\end{center}
\small{
{\bf Figure 2. Identifiability {\em tableaus} for Fed-batch reactor for ethanol production model: a) numerical initial conditions [1, 150, 0, 10] and  b) generic initial conditions.}

\subsubsection*{Case study 5: A linear biochemical pathway with 14 steps}

We perform structural identifiability analysis for the model (10) described in the main document. We consider the case with generic initial conditions and zero initial conditions except $x_1(0)=2$. The input file is the following: 
\vspace*{-0.35cm}

\begin{center}
\line(1,0){420}
{\small
\begin{verbatim}
             syms x1 x2 x3 x4 x5 x6 x7 x8 x9 x10 x11 x12 x13 x14
             syms x01 x02 x03 x04 x05 x06 x07 x08 x09 x010 x011 x012 x013 x014
             syms vm km p1 p2 p3 p4 p5 p6 p7 p8 p9 p10 p11 p12 p13 p14
             disp('Number of derivatives'); 
             Nder=2
             disp('Number of states'); 
             Neq=14
             disp('Number of model parameters'); 
             Npar=16
             disp('Number of controls'); 
             Noc=1
             disp('Number of observables'); 
             Nobs=14
                X=[x1 x2 x3 x4 x5 x6 x7 x8 x9 x10 x11 x12 x13 x14];
             
             disp('Equations of the model')
                A1 = -vm*x1/(km+x1);
                A2 = p1*x1-p2*x2;
                A3 = p2*x2-p3*x3;
                A4 = p3*x3-p4*x4;
                A5 = p4*x4-p5*x5;
                A6 = p5*x5-p6*x6;
                A7 = p6*x6-p7*x7;
                A8 = p7*x7-p8*x8;
                A9 = p8*x8-p9*x9;
                A10 = p9*x9-p10*x10;
                A11 = p10*x10-p11*x11;
                A12 = p11*x11-p12*x12;
                A13 = p12*x12-p13*x13;
                A14 = p13*x13-p14*x14;
                F=[A1 A2 A3 A4 A5 A6 A7 A8 A9 A10 A11 A12 A13 A14];
            
             disp('Controls')
                g1=p1;g2=0;g3=0;g4=0;g5=0;g6=0;g7=0;g8=0;g9=0;g10=0;
                g11=0; g12=0;g13=0;g14=0;
                G=[g1 g2 g3 g4 g5 g6 g7 g8 g9 g10 g11 g12 g13 g14];
             
             disp('Observables')
                h1=x1;h2=x2;h3=x3;h4=x4;h5=x5;h6=x6;h7=x7;h8=x8;h9=x9;h10=x10;
                h11=x11;h12=x12;h13=x13;h14=x14;
                H=[h1 h2 h3 h4 h5 h6 h7 h8 h9 h10 h11 h12 h13 h14];
             
             disp('Initial conditions')
                IC=[x01 x02 x03 x04 x05 x06 x07 x08 x09 x010 x011 x012 x013 x014];
             
             disp('Parameters')
                Par=[vm km p1 p2 p3 p4 p5 p6 p7 p8 p9 p10 p11 p12 p13 p14]; 

            GenSSI_generating_series(F,G,Noc,Neq,Nder,Nobs,H,X,IC,Par,result_folder);
\end{verbatim}
}
\vspace*{-0.35cm}
\line(1,0){420}
\end{center}

The model is structurally globally identifiable, as all the parameters have unique solution. The same results are obtained when using numeric initial conditions. In Figure 3 the identifiability \textit{tableaus} is presented, for both cases we get the same representation. 

\begin{center}
\epsffile{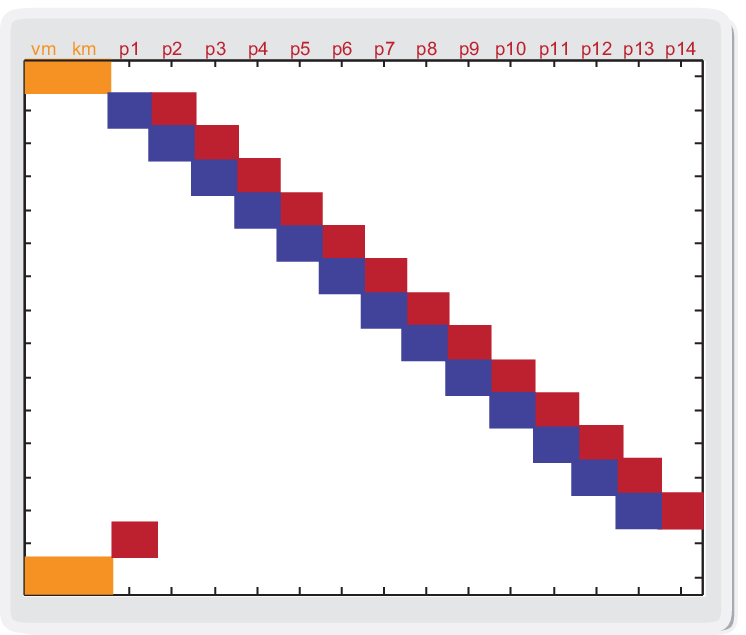} 
\end{center}
\small{
{\bf Figure 3. Identifiability {\em tableaus} for the linear biochemical pathway with 14 steps:} In red, direct globally identifiable parameters and in orange 2nd level globally identifiable parameters.}

\newpage
\subsubsection*{Case study 6: Six-gene regulatory network}

The model as proposed in DREAM6 challenge is described by a set of 12 ordinary differential equations and 35 parameters as presented in the main text. To perform the structural identifiability analysis we used a polynomial form of the original model. The input file for GenSSI toolbox is the following:
\vspace*{-0.35cm}
\begin{center}
\line(1,0){420}
{\small
\begin{verbatim}
       syms mr1 p1 mr2 p2 mr3 p3 mr4 p4 mr5 p5 mr6 p6 x13 x14 x15 x16 x17 x18 x19 x20
       syms pdr pst1 pst2 pst3 pst4 pst5 pst6 rst1 rst2 rst3 rst4 rst5 rst6 ...
            k1 h1 k2 h2 k3 h3 k4 h4 k5 h5 k6 h6 k7 h7 k8 h8
       disp('Number of derivatives'); 
       Nder=2
       disp('Number of states'); 
       Neq=20
       disp('Number of model parameters'); Npar=29
       disp('Number of controls'); Noc=0
       disp('Number of observables'); Nobs=20
       X=[mr1 p1 mr2 p2 mr3 p3 mr4 p4 mr5 p5 mr6 p6 x13 x14 x15 x16 x17 x18 x19 x20];
       
       disp('Equations of the model')
           A1 = (pst1) - mr1;
           A2 = rst1*mr1 - pdr*p1;
           A3 = (pst2*(x13/(1 + x13))*(1.0/(1 + x16))) - mr2;
           A4 = rst2*mr2 - pdr*p2;
           A5 = (pst3*(x15/(1 + x15))*(1.0/(1 + x19))) - mr3;
           A6 = rst3*mr3 - pdr*p3;
           A7 = (pst4*(x14/(1 + x14))*(1.0/(1 + x17))) - mr4;
           A8 = rst4*mr4 - pdr*p4;
           A9 = (pst5*(1.0/(1 + x18))) - mr5;
           A10 = rst5*mr5 - pdr*p5;
           A11 = (pst6*(1.0/(1 + x20))) - mr6;
           A12 = rst6*mr6 - pdr*p6;
           A13 = h2*x13/p1;
           A14 = h1*x14/p1;
           A15 = h3*x15/p1;
           A16 = h5*x16/p6;
           A17 = h8*x17/p5;
           A18 = h6*x18/p4;
           A19 = h4*x19/p2;
           A20 = h7*x20/p4;
           F=[A1 A2 A3 A4 A5 A6 A7 A8 A9 A10 A11 A12 A13 A14 A15 A16 A17 A18 A19 A20];
        
       disp('Controls')
           g1=p1;g1=0;g2=0;g3=0;g4=0;g5=0;g6=0;g7=0;g8=0;g9=0;g10=0; 
           g11=0;g12=0;g13=0;g14=0;g15=0;g16=0;g17=0;g18=0;g19=0;g20=0;
           G=[g1 g2 g3 g4 g5 g6 g7 g8 g9 g10 g11 g12 g13 g14 g15 g16 g17 g18 g19 g20];
          
      disp('Observables')
           h1=mr1;h2=p1;h3=mr2;h4=p2;h5=mr3;h6=p3;h7=mr4;h8=p4;h9=mr5;h10=p5;
           h11=mr6;h12=p6;h13=x13;h14=x14;h15=x15;h16=x16;h17=x17;h18=x18;h19=x19;h20=x20;
           H=[h1 h2 h3 h4 h5 h6 h7 h8 h9 h10 h11 h12 h13 h14 h15 h16 h17 h18 h19 h20];
     
      disp('Initial conditions')
           IC=[0 1 0 1 0 1 0 1 0 1 0 1 (1/k2)^h2 (1/k1)^h1 (1/k3)^h3 (1/k5)^h5 ...
              (1/k8)^h8 (1/k6)^h6 (1/k4)^h4 (1/k7)^h7];
          
      disp('Parameters')
           Par=[pdr pst1 pst2 pst3 pst4 pst5 pst6 rst1 rst2 rst3 rst4 rst5 rst6 ...
                 k1 h1 k2 h2 k3 h3 k4 h4 k5 h5 k6 h6 k7 h7 k8 h8]; 

      GenSSI_generating_series(F,G,Noc,Neq,Nder,Nobs,H,X,IC,Par,result_folder);
\end{verbatim}
}
\vspace*{-0.35cm}
\line(1,0){420}
\end{center}        

The model is structurally globally identifiable, as all the parameters have unique solution. The same results are obtained when using numeric initial conditions. In Figure 4 the identifiability \textit{tableaus} is presented, for both cases we get the same representation. 

\begin{center}
\epsffile{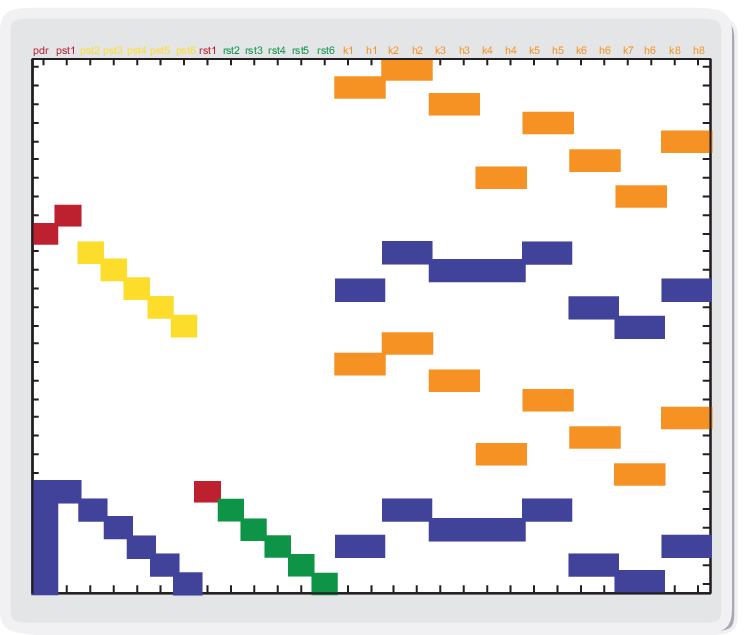} \\
\end{center}
\small{
{\bf Figure 4. Identifiability {\em tableau} for the linear biochemical pathway with 14 steps:} In red, direct globally identifiable parameters; in orange, yellow and green, 2nd to 4th level globally identifiable parameters.}

\newpage

\subsection*{The geometry of sloppiness with a simple example}

In the main text we showed how two different experimental schemes with different identifiability results may present similar sloppiness. We argue that it is possible to modify the confidence volume of the confidence hyper-ellipsoid without modifying the ratio between the maximum and the minimum semi-axis, i.e. the sloppiness. 

This is quite trivial if we consider a two dimensional example:

\begin{center}
\epsffile{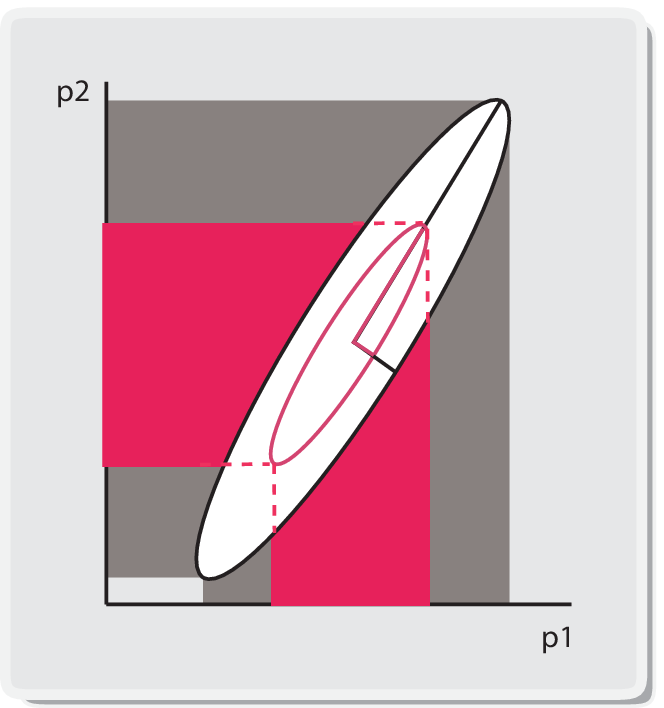} \\
\end{center}
\small{
{\bf Figure 5. Analysis of sloppiness from a geometric point of view.} Figure represents two confidence ellipses for an arbitrary example with two-parameters. The semi-axes of the larger ellipse are twice the semi-axes of the smaller ellipse. Confidence on the parameter values is better for the smaller ellipse whereas sloppiness is the same in both cases.}

Let's consider a simple dynamic model that we can treat symbolically. The system could describe a linear biochemical network and it is represented by two differential equations as follows:

\begin{equation}\label{cs7}
\begin{array}{ll}
\dot{x}_1=-p_1x_1, \\
\dot{x}_2=p_1x_1-p_2x_2,
\end{array}
\end{equation}

\noindent where $x_1$ and $x_2$ represent the concentrations of the two reaction components with the initial conditions $x_1(0)=c_1$ and $x_2(0)=0.$ If we take a single measurement at final time of both states, the sloppiness of the model can be computed as follows:

\begin{equation}
{\cal C}_F=\frac{e^{-2 ln(p_2) t} ln(p_1)^4 \left(e^{ln(p_2) t} ln(p_2)^2+e^{ln(p_1) t} ln(p_1) (ln(p_2)
   ln(p_1) t+ln(p_1)-ln(p_2) (ln(p_2) t+2))\right)^2}{(ln(p_1)-ln(p_2))^4 ln(p_2)^4 (ln(p_1) t+1)^2}
\end{equation}

\noindent Note that sloppiness depends on the values of the parameters and the sampling time, but it is independent of the initial condition of the experiments, whereas the eigenvalues of the Fisher information matrix, thus the confidence on the parameters, do. Figure 6 presents some numerical examples.

\begin{center}
\epsffile{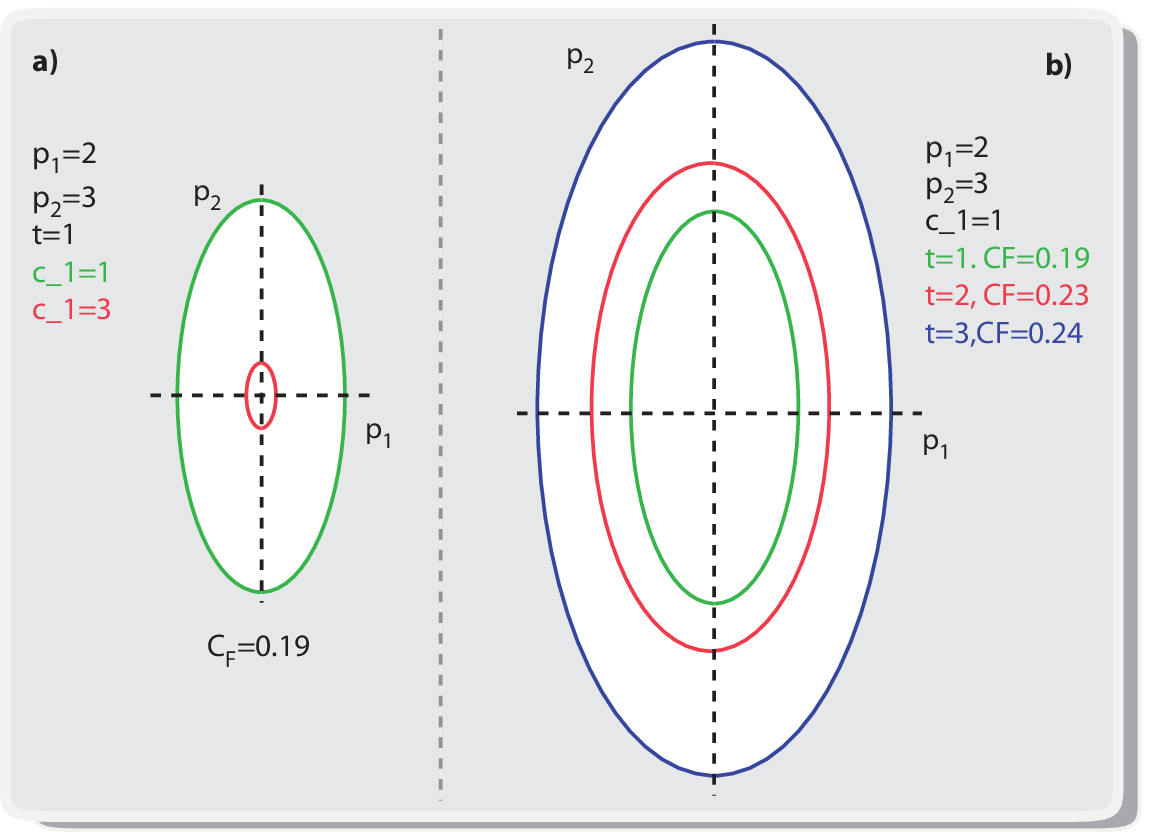} \\
\end{center}
\small{
{\bf Figure 6. Effect of the experimental setup in the sloppiness of a simple dynamic model.} Figure a) presents the effect of the initial conditions on the confidence ellipse, note that sloppiness is the same in both cases; Figure b) presents the confidence ellipses obtained when measuring at different sampling times. Sloppiness is practically the same in all cases. Results reveal that, since sensitivities decrease with time, the latest we measure the worse to identify the parameters.}

Numerical examples in Figure 6 do not correspond to a sloppy scenario. Is there any possibility to obtain a sloppy scenario with this example? In fact, if parameters have different orders of magnitude, we observe that the sensitivities with respect to parameters vary with implications on the sloppiness of the system. 
 
\begin{center}
\epsffile{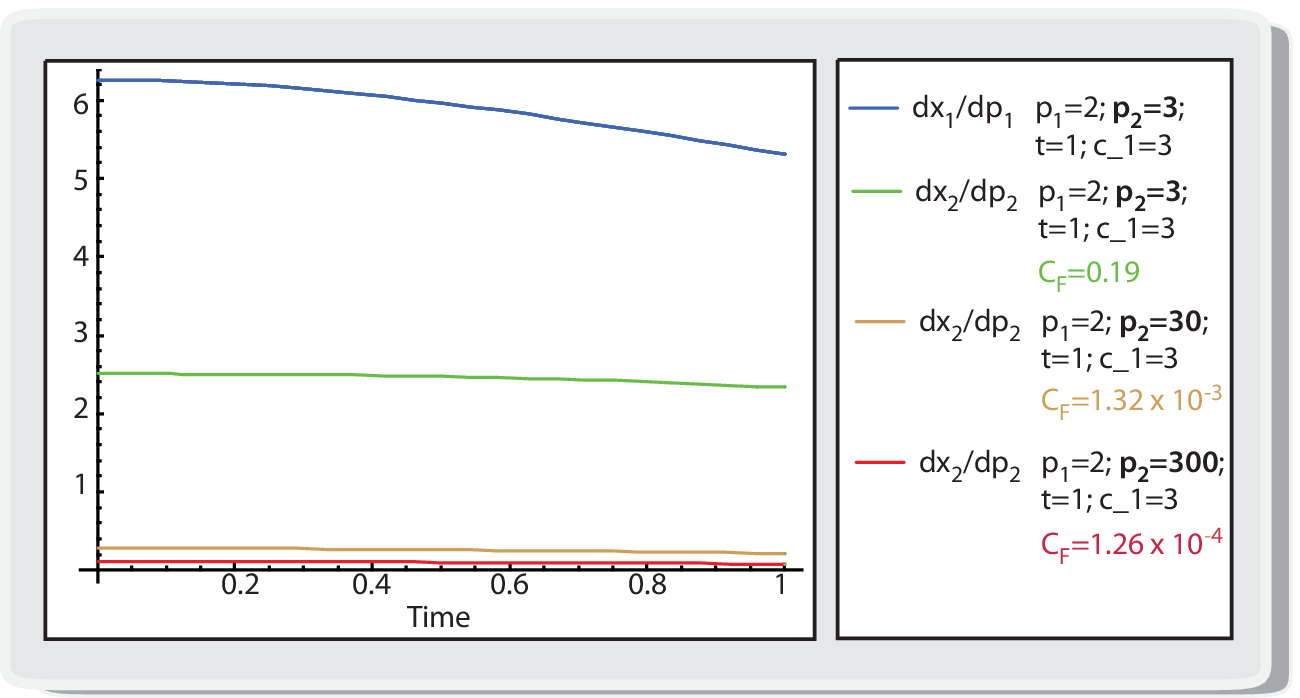} \\
\end{center}
\small{
{\bf Figure 7. Parametric sensitivities for the simple dynamic model.} Figure presents the evolution of the parametric sensitivities of states $x_1$ and $x_2$ for different values of the parameters. $\partial x_1/\partial p_1$ is independent of $p_2$ therefore all sensitivities coincide. The effects of modifying $p_2$ on the sensitivities of $x_2$ imply that the model slowly becomes sloppy.}

It should be noted that provided it is possible to incorporate $x_2$ at the beginning of the process ($x_2(0)\neq 0$) or it is possible to add a certain amount of $x_1$ to the system, it is possible to reduce sloppiness.

\end{document}